\newcommand\apjcls{1}
\newcommand\aastexcls{2}
\newcommand\othercls{3}

\newcommand\papercls{\aastexcls}
\documentclass[iop,
numberedappendix,
twocolappendix,
apj]{emulateapj}

\if\papercls \apjcls
\usepackage{apjfonts}
\else\if\papercls \othercls
\usepackage{epsfig}
\fi\fi
\usepackage{ifthen}
\usepackage{natbib}
\usepackage{amssymb, amsmath}
\usepackage{cancel} % for \cancel{}, \bcancel{}, \xcancel{}
\usepackage{booktabs} % nice tables
\usepackage{appendix}
\usepackage{etoolbox}
\usepackage[T1]{fontenc}
\usepackage{paralist}
\usepackage{commath} % some cool math features such as alignment of ":="

\usepackage[utf8]{inputenc}

\if\papercls \apjcls
\newcommand\aas{\ref@jnl{AAS Meeting Abstracts}}% *** added by jh
\newcommand\dps{\ref@jnl{AAS/DPS Meeting Abstracts}}% *** added by jh
\newcommand\maps{\ref@jnl{MAPS}}% *** added by jh
\else\if\papercls \othercls
\usepackage{astjnlabbrev-jh}
\fi\fi

\usepackage[%pdftex,      % hyper-references for pdflatex
bookmarks=true,           %%% generate bookmarks ...
bookmarksnumbered=true,   %%% ... with numbers
colorlinks=true,          % links are colored
citecolor=blue,           % color for \cite{...} links
linkcolor=blue,           % color for \ref{...} links
menucolor=blue,           % color for Acrobat menu buttons
urlcolor=blue,            % color for \url{...} links
linkbordercolor={0 0 1},  %%% blue frames around links
pdfborder={0 0 1},
frenchlinks=true]{hyperref}
\if\papercls \othercls

\else

\fi

\providecommand{\adsurl}[1]{\href{#1}{ADS}}

\makeatletter
\patchcmd{\NAT@citex}
  {\@citea\NAT@hyper@{%
     \NAT@nmfmt{\NAT@nm}%
     \hyper@natlinkbreak{\NAT@aysep\NAT@spacechar}{\@citeb\@extra@b@citeb}%
     \NAT@date}}
  {\@citea\NAT@nmfmt{\NAT@nm}%
   \NAT@aysep\NAT@spacechar\NAT@hyper@{\NAT@date}}{}{}

\patchcmd{\NAT@citex}
  {\@citea\NAT@hyper@{%
     \NAT@nmfmt{\NAT@nm}%
     \hyper@natlinkbreak{\NAT@spacechar\NAT@@open\if*#1*\else#1\NAT@spacechar\fi}%
       {\@citeb\@extra@b@citeb}%
     \NAT@date}}
  {\@citea\NAT@nmfmt{\NAT@nm}%
   \NAT@spacechar\NAT@@open\if*#1*\else#1\NAT@spacechar\fi\NAT@hyper@{\NAT@date}}
  {}{}
\makeatother

\makeatletter
\DeclareRobustCommand{\lowcase}[1]{\@lowcase#1\@nil}
\def\@lowcase#1\@nil{\if\relax#1\relax\else\MakeLowercase{#1}\fi}
\makeatother

\DeclareSymbolFont{UPM}{U}{eur}{m}{n}
\DeclareMathSymbol{\umu}{0}{UPM}{"16}
\let\oldumu=\umu
\renewcommand\umu{\ifmmode\oldumu\else\math{\oldumu}\fi}
\newcommand\micro{\umu}
\if\papercls \othercls
\newcommand\micron{\micro m}
\else
\renewcommand\micron{\micro m}
\fi

\let\oldsim=\sim
\renewcommand\sim{\ifmmode\oldsim\else\math{\oldsim}\fi}
\let\oldpm=\pm
\renewcommand\pm{\ifmmode\oldpm\else\math{\oldpm}\fi}
\newcommand\by{\ifmmode\times\else\math{\times}\fi}
\newcommand\tablebox[1]{\begin{tabular}[t]{@{}l@{}}#1\end{tabular}}
\newbox{\wdbox}
\renewcommand\c{\setbox\wdbox=\hbox{,}\hspace{\wd\wdbox}}
\renewcommand\i{\setbox\wdbox=\hbox{i}\hspace{\wd\wdbox}}

\newcount\timect
\newcount\hourct
\newcount\minct
\newcommand\now{\timect=\time \divide\timect by 60
         \hourct=\timect \multiply\hourct by 60
         \minct=\time \advance\minct by -\hourct
         \number\timect:\ifnum \minct < 10 0\fi\number\minct}

\catcode`@=11

\newcommand\comment[1]{}

\newcommand\commenton{\catcode`\%=14}

\renewcommand\math[1]{$#1$}
\newcommand\mathshifton{\catcode`\$=3}

\let\atab=&
\newcommand\atabon{\catcode`\&=4}

\let\oldmsp=\sp
\let\oldmsb=\sb
\def\sp#1{\ifmmode
           \oldmsp{#1}%
         \else\strut\raise.85ex\hbox{\scriptsize #1}\fi}
\def\sb#1{\ifmmode
           \oldmsb{#1}%
         \else\strut\raise-.54ex\hbox{\scriptsize #1}\fi}
\newbox\@sp
\newbox\@sb
\def\sbp#1#2{\ifmmode%
           \oldmsb{#1}\oldmsp{#2}%
         \else
           \setbox\@sb=\hbox{\sb{#1}}%
           \setbox\@sp=\hbox{\sp{#2}}%
           \rlap{\copy\@sb}\copy\@sp
           \ifdim \wd\@sb >\wd\@sp
             \hskip -\wd\@sp \hskip \wd\@sb
           \fi
        \fi}
\def\msp#1{\ifmmode
           \oldmsp{#1}
         \else \math{\oldmsp{#1}}\fi}
\def\msb#1{\ifmmode
           \oldmsb{#1}
         \else \math{\oldmsb{#1}}\fi}

\def\supon{\catcode`\^=7}

\def\subon{\catcode`\_=8}

\def\supsubon{\supon \subon}

\newcommand\actcharon{\catcode`\~=13}

\newcommand\paramon{\catcode`\#=6}

\comment{And now to turn us totally on and off...}

\newcommand\reservedcharson{ \commenton  \mathshifton  \atabon  \supsubon 
                             \actcharon  \paramon}

\catcode`@=12
\reservedcharson

\if\papercls \apjcls

\else

\fi

\if\papercls \othercls
\else
  \newcommand\inpress{y}
  \if\inpress y
    \received{2018 November 6}
    \revised{2019 January 17}
    \accepted{2019 February 9}
    \if\papercls \apjcls
    \slugcomment{}
    \fi
  \else
  \slugcomment{\tablebox{In preparation for {\em ApJ}. DRAFT of {\today}.}}
  \fi
\fi

\newcommand\chisq{\ifmmode{\chi\sp{2}}\else\math{\chi\sp{2}}\fi}
\newcommand\redchisq{\ifmmode{ \chi\sp{2}\sb{\rm red}}
                    \else\math{\chi\sp{2}\sb{\rm red}}\fi}
\newcommand\Teq{\ifmmode{T\sb{\rm eq}}\else$T$\sb{eq}\fi}
\newcommand\mjup{\ifmmode{M\sb{\rm Jup}}\else$M$\sb{Jup}\fi}
\newcommand\rjup{\ifmmode{R\sb{\rm Jup}}\else$R$\sb{Jup}\fi}
\newcommand\msun{\ifmmode{M\sb{\odot}}\else$M\sb{\odot}$\fi}
\newcommand\rsun{\ifmmode{R\sb{\odot}}\else$R\sb{\odot}$\fi}
\newcommand\mearth{\ifmmode{M\sb{\oplus}}\else$M\sb{\oplus}$\fi}
\newcommand\rearth{\ifmmode{R\sb{\oplus}}\else$R\sb{\oplus}$\fi}
\shorttitle{Pebbel Flux regulated Planetesimal formation}
\shortauthors{Lenz {\em et al.}}

\newcommand{\mstar}{M_{\star}} % star mass
\newcommand{\rchar}{{r_\mathrm{c}}} % "characteristic radius"
\newcommand{\vispow}{\gamma} % viscosity power-law index
\newcommand{\sigchar}{\Sigma_\mathrm{c}} % normalization constant
\newcommand{\stokes}{\mathrm{St}} % Stokes number
\newcommand{\stmin}{\stokes_\mathrm{min}} % minimum Stokes number
\newcommand{\stmax}{\stokes_\mathrm{max}} % maximum Stokes number
\newcommand{\stfrag}{\stokes_\mathrm{frag}} % fragmentation limit
\newcommand{\stdrift}{\stokes_\mathrm{drift}} % drift limit
\newcommand{\vfrag}{v_\mathrm{f}} % fragmentation threshold speed
\newcommand{\Mdisk}{M_{\mathrm{disk}}} % mass of the disk
\newcommand{\aturb}{\alpha_\mathrm{t}} % alpha turbulence parameter
\newcommand{\press}{P} % pressure
\newcommand{\turbvisc}{\nu} % turbulent viscosity
\newcommand{\surfp}{\Sigma_\mathrm{p}} % column density planetesimals
\newcommand{\surfpeb}{\Sigma_\mathrm{peb}} % pebble column density
\newcommand{\surfgas}{\Sigma_\mathrm{g}} % column density planetesimals
\newcommand{\dotsurfp}{\dot{\Sigma}_\mathrm{p}} % planetesimal column density rate
\newcommand{\pebflux}{\dot{M}_\mathrm{peb}} % radial pebble flux or, more precisely, the radial pebble mass rate
\newcommand{\totflux}{\dot{M}_\mathrm{tot}} % radial total particle flux
\newcommand{\frd}{\xi} % fragment distribution power-law index
\newcommand{\Mcr}{\dot{M}_\mathrm{cr}} % critical accretion rate of particles with St_min<St<St_max
\newcommand{\lc}{l_\mathrm{c}} % critical lenght scale
\newcommand{\peff}{\varepsilon} % Trapping and planetesimal formation efficiency
\newcommand{\trapdist}{d} % trap distance
\newcommand{\tlife}{\tau_\mathrm{l}} % trap lifetime
\newcommand{\AU}{\mathrm{au}} % astronomical unit
\newcommand{\yr}{\mathrm{yr}} % years
\newcommand{\K}{\mathrm{K}} % Kelvin
\newcommand{\dustrad}{a} % radius of spherical particles
\newcommand{\dustmass}{m_\mathrm{d}} % mass of dust particles
\newcommand{\stav}{\overline{\stokes}} % flux averaged Stokes number contributing to planetesimals
\newcommand{\dustradav}{\overline{\dustrad}} % flux averaged Stokes number contributing to planetesimals
\newcommand{\vdrift}{v_\mathrm{drift}} % drift velocity of particle
\newcommand{\dustsurf}{\Sigma_{\mathrm{d}}} % dust column density
\newcommand{\nm}{n_{\mathrm{m}}} % number density per unit mass in Smoluchowski Equation
\newcommand{\numdens}{n} % number density
\newcommand{\nsize}{n_{\dustrad}} % number density per radius of solid particles
\newcommand{\mptes}{m_\mathrm{p}} % planetesimal mass
\newcommand{\mdust}{m} % mass of particles (dust grains)
\newcommand{\msol}{\mdust} % mass of solid particles
\newcommand{\dice}{f_\mathrm{ice}} % dust ice parameter
\newcommand{\hscale}{h_\mathrm{g}} % gas pressure scale height
\newcommand{\hscdust}{h_\mathrm{d}} % dust/particle scale height
\newcommand{\Okepl}{\Omega} % Keplerian angular velocity
\newcommand{\Grav}{G} % Newton's gravitational constant
\newcommand{\vdisp}{{\delta v}} % velocity dispersion
\newcommand{\cs}{c_\mathrm{s}} % isothermal sound speed
\newcommand{\expT}{q} % temperature power-law index

\newcommand{\mdens}{\rho} % mass density
\newcommand{\rhoint}{\mdens_\mathrm{s}} % solid density, i.e., internal density
\newcommand{\dtog}{Z} % dust-to-gas ratio
\newcommand{\rhog}{{\mdens_{\mathrm{g}}} } % gas density
\newcommand{\rhod}{ {\mdens_{\mathrm{d}}} } % dust density
\newcommand{\sound}{c_{\mathrm{s}}} % sound speed
\newcommand{\vkepl}{v_{\mathrm{K}}} % Kepler speed (scalar)
\newcommand{\torb}{t_\mathrm{orb}} % orbit time
\newcommand{\vth}{v_\mathrm{th}} % mean thermal speed for Maxwell-Boltzmann distribution
\newcommand{\vdustk}[1]{v_\mathrm{#1}} % dust velocity component
\newcommand{\vgask}[1]{u_\mathrm{#1}} % gas velocity component
\newcommand{\diffd}{D_\mathrm{d}} % dust diffusivity
\newcommand{\diffg}{D_\mathrm{g}} % dust diffusivity

\newcommand{\fprob}{p_\mathrm{f}} % fragmentation probability
\newcommand{\cprob}{p_\mathrm{c}} % coagulation probability
\newcommand{\coagkern}{C} % coagulation kernel
\newcommand{\fragkern}{F} % fragmentation kernel
\newcommand{\fragdistr}{\mathcal{D}} % distributions after a collision, _{\mathrm{frag}}
\newcommand{\rice}{r_\mathrm{ice}} % radius of the ice line
\newcommand{\vrelcoll}{\Delta{v}} % Relativgeschwindigkeit bei Kollision von festen Partikeln
\newcommand{\crossseccoll}{\sigma} % cross section of the collision
\newcommand{\DD}[1]{\delta\left(#1\right)} % Dirac delta function. _{\mathrm{D}}

\renewcommand{\pd}[2]{\frac{\partial {#1}}{\partial {#2}}}
\newcommand{\dfracpd}[2]{\dfrac{\partial {#1}}{\partial {#2}}} % \pd{}{} with \dfrac

\newcommand{\ie}{{i.\,e.}}
\newcommand{\Ie}{{I.\,e.}}
\newcommand{\eg}{{e.\,g.}}

\shorttitle{Pebble Flux regulated planetesimal formation}
\shortauthors{Lenz et al.\ }

\begin{document}


\title{Planetesimal Population Synthesis:\\ Pebble Flux-regulated Planetesimal Formation}

% Trapping is the parameter by which the flux is converted.
%\chris{Trapping} 


\author{Christian~T.~Lenz\altaffilmark{1,}$^{\star}$,
Hubert~Klahr\altaffilmark{1},
and Tilman~Birnstiel\altaffilmark{2}
}

\affil{\sp{1} Max Planck Institute for Astronomy, K{\"o}nigstuhl 17, D-69117 Heidelberg, Germany, \href{mailto:lenz@mpia.de}{lenz@mpia.de},
\href{mailto:klahr@mpia.de}{klahr@mpia.de}
\\
       \sp{2} University Observatory, Faculty of Physics, Ludwig-Maximilians-Universit{\"a}t M{\"u}nchen, Scheinerstr. 1, D-81679 Munich, Germany\\
    %   \sp{3} Harvard-Smithsonian Center for Astrophysics, 60 Garden Street, 02138 Cambridge, MA, USA\\
       $^\star$ Member of the International Max Planck Research School for Astronomy and Cosmic Physics at the Heidelberg University
       }

% \author{Christian T. Lenz\altaffilmark{1}, Hubert Klahr\altaffilmark{1}, \& Tilman Birnstiel\altaffilmark{2}}
% \email{lenz@mpia.de, klahr@mpia.de}
%\email{VERSION: \today}
%\email{STATUS: draft}

% \altaffiltext{1}{Max-Planck-Institut f\"ur Astronomie, K\"onigstuhl
%                  17, 69117, Heidelberg, Germany}
% \altaffiltext{2}{LMU, Munich}



\begin{abstract}
We propose an expression for a local planetesimal formation rate
proportional to the instantaneous radial pebble flux. The result---a radial planetesimal 
distribution---can be used as initial condition to study the formation of planetary embryos. 
We follow the idea that one needs particle traps to locally enhance the dust-to-gas ratio
sufficiently such that particle gas interactions can no longer prevent planetesimal
formation on small scales. The location of these traps can emerge everywhere in the disk.
Their occurrence and lifetime is subject of ongoing research, thus here 
they are implemented via free parameters. 
This enables us to study the influence of the disk properties on the formation of planetesimals, 
predicting their time dependent formation rates
and location of primary pebble accretion.
We show that large $\alpha$-values of $0.01$ (strong turbulence) prevent the 
formation of planetesimals in the inner part of the disk, arguing for lower 
values of around $0.001$ (moderate turbulence), at which planetesimals form quickly at all
places where they 
are needed for proto-planets. Planetesimals form as soon as dust has grown to pebbles 
($\sim\mathrm{mm}$ to $\mathrm{dm}$) and the pebble flux reaches a critical value, 
which is after a few thousand years at $2-3\,\AU$ and after a few hundred thousand years at $20-30\,\AU$.
Planetesimal formation lasts until the pebble supply has decreased below a critical value. 
The final spatial planetesimal distribution is steeper compared to the initial dust and 
gas distribution which helps to explain the discrepancy between the minimum mass solar 
nebula and viscous accretion disks.
\end{abstract}


\noindent
%$^\dagger$UCO/Lick Observatory Bulletin, No..........


% C.T.L.: https://orcid.org/0000-0002-4202-2457
% T.B.:   https://orcid.org/0000-0002-1899-8783
% H.K.:   https://orcid.org/0000-0002-8227-5467

\section{Introduction}
The original ideas about planet formation---valid for terrestrial planets as well as gas- and ice-giants---assume that pebble-sized material, which has grown from tiny dust and ice grains via successive collisions and 
sedimented towards the midplane of the disk, directly collapses into $\sim10-100\,\mathrm{km}$ sized planetesimals \citep{Safronov1969,GoldreichWard1973}.
By definition planetesimals are bodies massive enough to be bound by, and able to accrete further solid material via, gravitational attraction. 
A critical size or mass for boulders to be called planetesimal can be 
estimated by requiring that pebbles settle toward the planetesimal with a higher 
velocity compared with the typical headwind the planetesimal experiences in the 
gaseous environment \citep{OrmelKlahr2010}. The precise mass depends on the 
location in the disk and the size of the pebbles. 
The fact that their binding force is gravity makes it plausible gravity could also be 
involved in their formation in the first place. Nevertheless, \cite{weidenschilling1995} showed that 
\cite{Safronov1969} and \cite{GoldreichWard1973} neglected the effect of friction between particles and gas, 
and the associated momentum transfer from the solids to the gas.
Sedimentation to the midplane leads to dust-to-gas ratios of the order of unity before the critical density for self-gravity among the pebbles is reached. 
But starting from dust-to-gas ratios of unity the drag from the particles accelerates the gas disk, which is usually pressure supported and sub-Keplerian, to the Keplerian speed. 

The feedback from the dust onto the gas modifies the average azimuthal velocity of
the gas which leads to resonant drag instabilities \citep{SquireHopkins2018b,SquireHopkins2018a}, for example the streaming instability \citep{YoudinGoodman2005}.
The resulting turbulence will
then prevent further vertical sedimentation and, ultimately, gravitational collapse of pebble clouds to planetesimals, 
because the necessary critical density in the pebble cloud cannot be reached. Originally this  
effect was identified as a Kelvin-Helmholtz instability \citep{sekiya1998, JohansenHenningKlahr2006}, yet in modern 3D simulations \citep{johansen2009} one finds streaming instability to dominate the dynamical evolution.

After this important 
discovery, research was focused on the growth from 
grains to planetesimals via sticking collisions. Yet this model was also facing several 
problems like the drift and fragmentation 
% When the relative velocities between grains are too large, they fragment. Since the relative velocity between 
% grains is typically rising with size, there is a size at which particles mostly fragment. This is called the fragmentation barrier. 
% On the other hand, if particles drift faster than they can grow, they will be removed before potentially growing to larger sizes. This is called the drift barrier. 
% But growth may also be limited by the so-called bouncing barrier \citep{guttler2010,zsom2010} or due 
% to charging of aggregates \citep{okuzumi2009} ---
% neither is considered in this paper.
barriers, because either the drift time becomes 
shorter than the growth time (drift limit) \citep{KlahrBodenheimer2006,Birnstiel2012} or 
the turbulence in the nebula induces collision speeds among the solids that lead 
to bouncing \citep{guttler2010,zsom2010} and fragmentation \citep{BlumMuench1993,BlumWurm2008,GundlachBlum2014}. Aggregates charging can also act as a growth barrier \citep{okuzumi2009}. Neither bouncing, nor charging is considered in this paper.
In the early stages of circumstellar disks, micrometer-sized dust grows to pebble-size (typically $\sim\mathrm{mm-dm}$) via collisions. Particle growth models have been developed and improved
in the past
\citep{zvyagina1974,weidenschilling1980,DullemondDominik2005,brauer2008coagulation,birnstiel2010,windmark2012,krijt2016b,
krijt2016a,stammler2017}.
\cite{Birnstiel2012} have shown that the radial dynamics
are mostly determined by particles with sizes close to the growth barriers.

Following the path of 
fluffy particle growth \citep{okuzumi2012,kataoka2013}, it becomes feasible to form planetesimals, at least within $10\,\AU$ 
in the disks. Yet, \cite{blum2017} have shown that the resulting planetesimals would have 
too large a tensile strength to explain the properties of comets, which are believed to be remnants from the planetesimal formation episode in the solar nebula.

The idea of \cite{Safronov1969} and \cite{GoldreichWard1973} was extended by \cite{YoudinShu2002}, showing that this gravitational instability can develop despite Kelvin-Helmholtz instability for sufficiently high dust-to-gas column density ratios at least in a laminar gas disk.

The work by \cite{Johansen2006} started a revival of gravity assisted planetesimal formation, because it was the first work that incorporated turbulence generated via the magneto rotational instability [MRI] \citep{BalbusHawley1991}. 
They found that turbulence not only leads to diffusion and prevents sedimentation, but also causes solid 
particles to get trapped and concentrated in
non-laminar flow features reaching locally a critical Hill density to undergo collapse.
% non-laminar flow features like zonal flows. 
These were 
created as by-products of the magneto rotational instability. They were
able to show that the over-densities are in fact sufficient to trigger gravitational collapse. 

This idea is different 
from the gravitational collapse envisioned by \cite{cuzzi2008}, in which grains get stochasticaly concentrated locally between vortices for a short time. The vortices there are part of a Kolmogoroff cascade and only live for a short time.

In the paper by \cite{johansen2007nat}
% \cite{johansen2009zonal}
they found that a zonal flow 
feature arose in a box simulation of the MRI simulation, 
in which sufficiently many particles got trapped to trigger a gravitational collapse, without being diffused by the streaming instability. 
In test simulations in the same paper without MRI and thus no zonal flows, the streaming instability prevented the formation of planetesimals.
These simulations of planetesimal formation in zonal flows were expanded to higher resolutions in order to study the size of forming planetesimals in \cite{johansen2012}, and the properties of zonal flows were further studied in \cite{Dittrich2013} and \cite{BaiStone2014}. Zonal flows appear to be transient features in any accretion disk subject to magnetic instabilities \citep{johansen2009zonal}.
In the simulations of \cite{Dittrich2013} they found a typical separation between the zonal flows of 5 pressure scale heights and a lifetime of about 100 local orbits. Whereas the above mentioned simulations were ideal magneto-hydrodynamic (MHD) runs, 
non ideal MHD simulations, especially those including the Hall-term, also showed the phenomenon of creating zonal flows \citep{BaiStone2014,bethune2016}. 

In situations where magnetic effects are weak because of the typically low ionization of disks around young stars, dust absorbs
free electrons and the field-lines decouple from the gas. In this situation potentially hydrodynamical instabilities could drive turbulence and structure formation. For instance a radial temperature gradient leads to 
vertical shear in the disk, which is unstable for short thermal relaxation times \citep{urpin1998,arlt2004,nelson2013,StollKley2014,barker2015,StollKley2016,latter2017,MangerKlahr2018,LyraUmurhan2018,klahr2018}. This was demonstrated to trigger zonal flows, long lived vortices \citep{raettig2013,Raettig2015,MangerKlahr2018}, and instability based on radial buoyancy.
The same can happen in disks which are radially buoyant, but for longer thermal relaxation times. Then, convective overstability \citep{KlahrHubbard2014,lyra2014} and its non-linear version, Subcritical Baroclinic Instabilities \citep{KlahrBodenheimer2003,petersen2007a,petersen2007b,lesur2010,LyraKlahr2011}, can produce and sustain vortices.

\cite{Johansen2007} performed 2D simulations on the Kelvin-Helmholtz instability of sedimenting particles and found that while on average particles were still diffused out of the midplane, the turbulence also led to strong fluctuations. 

In order to form planetesimals grains have to circumvent the growth barriers. One possibility is a jump 
from a clump of condensed pebbles directly to planetesimals as first shown in the above mentioned shearing-box simulations by  
\cite{johansen2007nat} in the gravo-turbulent scenario. 

Also in the complete absence of turbulence in the disk, which seems unlikely given the observation of accretion in these systems, sedimentation of pebbles in disks with high enough dust to gas ratios, that is $2-3$ times solar metalicity, can lead to gravitational instability and planetesimal formation, regulated by the streaming instability as originally explored by \citet{YoudinShu2002} and since then demonstrated in a series of numerical simulations \citep{johansen2009, simon2016}.

Independent from the source of turbulence (S.I. or external) there may exist a typical 
planetesimal diameter regulated by particle diffusion \citep{KlahrSchreiber2015,andydiss}. 
Grain evolution models, including radial transport, leading to planetesimal formation, have been performed by various authors. 
\cite{brauer2008planetesimal} have shown that particles with a radius of $\sim10^2\,\mathrm{m}$ can 
be produced at the edges of regions from low turbulence (called the ``dead zone'') to higher turbulence levels. 
A particle velocity distribution can lead to ``lucky particles'' which are able to circumvent 
the growth barriers via low relative speed collisions \citep{windmark2012,drazkowska2014proceeding}. 
But this effect may be suppressed by lower breakup speeds \citep{BlumMuench1993,BlumWurm2008} within the (water) ice line, and in the outer disk by rapid removal due to radial drift as indicated by \cite{estrada2016}. We will discuss this in more detail in section~\ref{sec:lucky}. 
\cite{drazkowska2013} have shown that planetesimal formation can occur in a pressure bump at the inner edge of the dead zone via sweep-up growth. 
Another possibility would be that particles
become fluffy, and grow continuously to planetesimal size. This model will be further discussed in section~\ref{sec:fluffy}. 
But particles can also pile up due to back-reaction of particles onto gas and
evaporation and 
re-condensation at the water ice line \citep{drazkowska2016,drazkowska2017} which we discuss in section~\ref{sec:SImodel}.

We will follow a different path where a plethora of various 
instabilities can create turbulent structures in almost the entire disk either by hydrodynamical
\citep{PfeilKlahr2018} or magnetic instabilities in the non ideal regime \citep{bethune2016}. For a review see \cite{klahr2018}. Indeed observations find a lot of structure in protoplanetary disks, see {\eg} the Disk Substructures at High Angular Resolution Project \citep[DSHARP; \eg,][]{andrews2018,huang2018,dullemond2018}. Hence, one can assume that trap structures will 
appear everywhere in the disk. These can then trap the incoming particles, which
are concentrated to particle clouds collapsing to planetesimals.

This paper is structured as follows. In section~\ref{sec:rational_trapping_mode} we present 
a new model for the planetesimal formation rate. Section~\ref{sec:nummodel} summarizes the numerical disk 
model and explains how we implemented our parameterized planetesimal formation within it. 
In section \ref{results} we show the first results.
We discuss 
limitations of the model and compare with other planetesimal formation models in section~\ref{sec:discussion}. 
Particular attention should be paid to section~\ref{sec:SImodel}, where we explain that streaming instability is not mandatory to form planetesimals. 
We conclude in section~\ref{sec:concl}.

\section{Rational of Pebble Flux limited Planetesimal formation: Trapping Mode.}\label{sec:rational_trapping_mode}
In order to mimic planetesimal formation in our simulations, we use findings from fluid dynamical simulations which include 
particles. To keep it simple and feasible we use a probabilistic toy model based on parameterization. 
The important disk model parameters for this model are the lifetime of pebble traps $\tlife$ ({\eg} vortices or zonal flows) and 
their radial separation $\trapdist$. Strictly speaking, one doesn't necessarily need a pressure \emph{bump} with a local maximum. If the pressure profile is flattened a bit without a local maximum, this would slow down the particles since smaller pressure gradients lead to slower drift speeds \citep{whipple1972} and conservation of mass flux leads to denser particle accumulations. Instead, the density is 
increased which may lead to critical $\dustsurf/\surfgas$ values to trigger streaming instability. In this scenario, the 
$\trapdist$ parameter has to be interpreted as the distance between such flat pressure structures. Or in general, $\trapdist$ is 
the distance between structures leading to streaming instability conditions with sufficient density increase to reach Hill 
density. In our case we just assume that the flux concentrators are vortices or zonal flows with typical 
numerically measured radial distance $5\hscale$ \citep{Dittrich2013}.

% The Stokes number of particles determines their dynamics. 
The Stokes number $\stokes$ describes the aerodynamic behavior of particles surrounded by gas. It is the 
ratio of the friction time (timescale of coupling to the gas) and the dynamical timescale of the gas 
(here the inverse of the Keplerian frequency $1/\Okepl$). Particles with $\stokes\ll1$ are well coupled to the gas whereas $\stokes\gg1$ are decoupled from the gas. 
From the particle side, the parameters of our model are the Stokes number of the smallest ($\stmin$) and largest ($\stmax$)
particles that are able to participate in the streaming instability to facilitate gravitational collapse of the particle
heaps. And finally an efficiency parameter $\peff$ that defines how much of the actual radial mass flux in pebbles can be trapped \emph{and} 
transformed into planetesimals. Then one can convert the column density of drifting pebbles into column density of planetesimals 
with the following recipe:
\begin{align}
\label{eq:dotsurfp}
 \dotsurfp(r)=\frac{\peff}{\trapdist(r)}\frac{\pebflux}{2\pi r}, %\int_{\stmin}^{\stmax}\abs{\vdrift(r,\stokes)}\dustsurf(r,\stokes)\dif\stokes\;.
\end{align}
where we defined the mass flux of pebbles for a full circumference as
\begin{align}
    \label{eq:pebflux}
	\pebflux:=2\pi r\sum_{\stmin\leq\stokes\leq\stmax}\abs{\vdrift(r,\stokes)}\dustsurf(r,\stokes).
\end{align}
Here, $\vdrift$ is the radial velocity with which particles drift
and $\dustsurf(r,\stokes)$ is the column density in 
particles with Stokes number $\stokes$ within a given bin in particle size. The units of this column density are still $\mathrm{g/cm^2}$. The cylindrical radius, \ie, distance to the star in the midplane, is given by $r$. Eq.~\eqref{eq:pebflux} is written in discretized form, as we treat it in our simulations. 

The \emph{conversion length} over which 
pebbles are transformed into planetesimals is given by
\begin{align}
	\label{eq:conv_len}
	\ell:=\trapdist/\peff.
\end{align}
If, for example, there is only one pebble 
species with drift speed $\vdrift$ and column density $\Sigma_\mathrm{peb}$, we distribute the new planetesimal column density over the trap distance $\trapdist$. $\ell/\vdrift$ gives the timescale of 
conversion, thus $\vdrift/\ell\cdot\Sigma_\mathrm{peb}$ is the rate at which 
the transformation from $\Sigma_\mathrm{peb}$ to $\surfp$ occurs. If this is added up for all 
particle species, one obtains the column density formation rate of planetesimals as shown in Eq.~\eqref{eq:dotsurfp}. 
We barely resolve the expected radial trap structures with a few radial bins at maximum. 

% $\vdrift/\trapdist$ is the inverse of the drift time scale these particles need to drift from one trap to another. 
% Or in other words, since $\trapdist$ was found to also be the trap extend \citep{Dittrich2013}, particles are drifting through the trap on time scales of $\trapdist/\vdrift$.
% \Ie, $\vdrift/\trapdist\cdot\dustsurf$ is the rate at which the column density is provided for trapping and planetesimal formation. 
To give another explanation, $1/\trapdist$ is the 
radial trap density, meaning that the closer the traps are packed radially, the more planetesimals can be produced. 
% The pressure bumps are not resolved by the numerical grid since they typically have a size of a few gas pressure scale heights. 
For the flux, only $\vdrift$ without the gas velocity is considered to take the relative radial velocity between particles and gas. Hereby we assume that the spatial pressure structures (\eg, real pressure-bump traps like vortices) 
move radially with the gas of the smooth profile.

We assume theoretically infinitesimal structures as traps even though they are practically extended. Since we assume a non-zero $\aturb$ value everywhere anyway, we also assume that turbulent structures can be found everywhere.  
For our numerical radial grid, the grid cells are smaller than the trap distance parameter $\trapdist$. This 
ensures that we can 
distribute the mass of newborn planetesimals accordingly (probabilistic ansatz). If the resolution is too low, \ie, 
$\Delta x\gg\trapdist$, a ``distribution'' of this mass is not physical or one cell traps too much mass. Also, the condition of accumulated particles may not be 
treated correctly anymore because a bigger cell can also host more mass. At the same time, we are far from resolving the length scale on which planetesimal formation occurs. This length scale is typically on the order of $0.01\hscale$ \citep[see his Eq. (3.40)]{andydiss}, depending on particle diffusion on these small scales and on particle Stokes number. Even though we may be able to resolve trap structures of the order of a gas pressure scale height with only a few radial bins, it is not enough to reach a proper resolution of the substructure. The properties of the traps in terms of particle trapping and planetesimal formation is embedded in our efficiency parameter $\peff$. 

The material density is assumed to be $\rhoint = 1.2\,\mathrm{g/cm^3}$ for all solid
particles, according to asteroid data \citep[\eg,][]{carry2012}.
For simplicity we again assume particles to be spherical objects with constant
material density $\rhoint$, such that the representative planetesimal mass with a diameter of $100\,\mathrm{km}$ \citep{KlahrSchreiber2015,andydiss} is given
by
\begin{equation}
	\label{eq:mptes}
	\mptes = \frac{4\pi}{3} (50\, \mathrm{km})^3\rhoint\approx 6.28\cdot 10^{20}\,\mathrm{g}\approx 1.05\cdot 10^{-7}\mearth
\end{equation}
where $\mearth$ is the mass of the earth. The idea of planetesimals of diameter around $100\,\mathrm{km}$ is also supported by data
from our solar system \citep{morbidelli2009,delbo2017}.

Planetesimal formation will only occur if, within one (average) lifetime of a trap $\tlife$, enough mass can 
be accumulated to form at least one planetesimal, {\ie} if the following condition holds
\begin{align}
	\label{eq:cond_ptes}
	\peff\tlife\pebflux>\mptes.
\end{align}
Hence, there is a critical flux 
\begin{align}
	\Mcr:=\frac{\mptes}{\peff\tlife}
\end{align}
that must be reached to allow planetesimal formation. If condition \eqref{eq:cond_ptes} is fulfilled, we call the flux \emph{critical}, otherwise we call it 
\emph{sub-critical}. We give more detailed information on that criterion in Appendix~\ref{cond_detail}.

We further assume that the relative speed between the gravoturbulence triggering structures and the particles is the drift speed $\vdrift$. 
But in reality, e.g., zonal flows can have radial velocities which would change this relative velocity. As long as the radial speed of the pressure bump structure is much smaller than the drift speed, the error is small. 
In this paper, again
for simplicity, we assume that the relative speed is 
always given by $\vdrift$. Additionally, since we don't know how $\peff$ 
would change with $\stokes$ and $r$, we assume that it is a constant.
As long as condition~\eqref{eq:cond_ptes} is fulfilled, only the value of 
the conversion length $\ell$ matters.
That is, one can change the value 
of $\peff$ and $\trapdist$ by the same factor, leading to the same 
result for planetesimals. 
% But again, only if the condition for the onset of planetesimal formation is fulfilled. 
% As we will show, most of the 
% planetesimals are build roughly between $10^4$ and a few $10^5\,\yr$ where 
% which means that if an increase of $\peff$ doesn't lead to a 
% sub-critical flux, the same $\ell$ will lead to the same result.

\section{Numerical Model}\label{sec:nummodel}
\subsection{The disk and dust model}
\begin{table*}[t]
\caption{Parameters which are kept constant in this paper. Values of the disk mass and of $\rchar$ can be compared with the statistics in Fig.~3 of \cite{andrews2010}. Per size decade we use 27 grid points.} 
\label{tab:paraconst}
\centering
\begin{tabular}{p{.07\textwidth}llp{.05\textwidth}l} % p{} weist einer Spalte eine gewisse Breite zu, sodass innerhalb einer Zelle Zeilenumbrüche erfolgen können
 \toprule\toprule
 Symbol   	& Value         		& Meaning \\
 \midrule
 $\mstar(t_0)$	& $1\,\msun$ 			& Initial mass of the central star \\
 $\Mdisk$    	& $0.05\,\msun$ 		& Total initial disk mass (gas+particles)								\\
 $\rchar$    	& $35\,\AU$     		& Transition radius from power-law to exponential drop off for $\surfgas$, Eq.~\eqref{eq:gasini}  		\\
 $\trapdist$ 	& $5\,\hscale$    		& Pebble trap distance, Eq.~\eqref{eq:dotsurfp} 		\\
 $\tlife$ 	& $100\,\torb$ 			& Lifetime of pebble traps 								\\
 $\dice$	& $1/3$ for $T>170\,\K$	& Dust ice parameter, Eq.~\eqref{eq:dice} \\
 $\dtog(t_0)$ 	& $10^{-2}$			& Initial dust-to-gas column density ratio, $\dustsurf/\surfgas$      		\\
 $\vispow$ 	& $1$ 				& Power-law index in gas density profile, Eq.~\eqref{eq:gasini} \\
 $\expT$ 	& $0.4$ 			& Power-law index of temperature profile, Eq.~\eqref{eq:Tprofile} \\
 $\frd$ 	& $1.83$ 			& Power-law index for fragment distribution, Eq.~\eqref{eq:frdist} \\
 $\vfrag$ 	& $1-10\,\mathrm{m/s}$ 		& Fragmentation threshold speed, Eq.~\eqref{eq:vfragm-r} \\
 $\dustrad_0$ 	& $0.1\,\mathrm{\micron}$ 		& Smallest grain radius for all times \\
 $\dustrad_1$ 	& $1\,\mathrm{\micron}$ 			& Largest grain radius for initial condition \\
 $\numdens$ 	& $\propto\dustrad^{-3.5}$ 	& Initial number density distribution, Eq.~\eqref{eq:nini} \\
 $\rhoint$ 	& $1.2\,\mathrm{g/cm^3}$ 	& Internal density of all solids (dust, pebbles, planetesimals) \\
 $\mptes$ 	& $1.05\cdot 10^{-7}\,\mearth$ 	& Planetesimal mass of $100\,\mathrm{km}$ diameter planetesimals, Eq.~\eqref{eq:mptes} \\
 $\stmin$	& $10^{-2}$					& Minimum Stokes number participating in trapping \& planetesimal formation\\
 $\stmax$	& $10$						& Maximum Stokes number participating in trapping \& planetesimal formation\\
%  $t_0$		& $100\,\mathrm{yr}$		& initial time of simulations \\
 \bottomrule
 %\bottomrule
\end{tabular}
\end{table*}
In the following, we will summarize the disk and dust model of the dust and gas evolution code from \cite{birnstiel2010} in which we have implemented our planetesimal formation model. 

We assume that turbulence is described by an effective kinematic viscosity of \citep{Shakura1973}
\begin{equation}
	\label{eq:turbvisc}
	\turbvisc = \aturb\cs\hscale.
\end{equation}
The dimensionless parameter $\aturb$ describes the strength of turbulence, since it determines the turbulent velocities. 
Most likely, the turbulent velocities of the largest eddies is roughly given by $\sqrt{\aturb}\cs$ as argued by \cite{cuzzi2001}.

We assume further that the gas is vertically in hydrostatic balance \citep{weizsacker1948} and that consequently the particle and gas density follow the Gaussian profile \citep{Safronov1969,Pringle1981} % Weiz: sec. 3b; safr: page 25; pringle: sec. 3.3
\begin{equation}
	\label{eq:rhoverticalstructure}
 	\rho_i(r,z)=\rho_{i,0}(r)\cdot\exp{\left[-\frac{1}{2}\left(\frac{z}{h_i}\right)^2\right]},
\end{equation}
where $i$ is either dust/particles ($\mathrm{d}$) or gas ($\mathrm{g}$). To eliminate the $z$-dependence, we use column densities\footnote{In Eq.~\eqref{eq:surfgas} $i$ can also be planetesimals (p).}
\begin{equation}
 \label{eq:surfgas}
 \Sigma_i(r,t):=\int_{-\infty}^{\infty}\rho_i(r,z,t)\dif z.
\end{equation}
The mass density of the gas at the mid-plane ($z=0$) can then be expressed as
\begin{equation}
 \label{eq:rhozerogas}
 \rho_{i,0}(r)=\frac{1}{\sqrt{2\pi}}\frac{\Sigma_i(r)}{h_i}.
\end{equation}
The gas pressure scale height is given by
\begin{equation}
	\hscale = \sound/\Okepl
\end{equation}
and the dust/particle scale height for $\stokes<1$ by \citep{Dubrulle1995} 
\begin{equation}
	\label{eq:hscdust}
	\hscdust=\left(1+\dfrac{\stokes}{\aturb}\right)^{-1/2}\hscale.
\end{equation}
Here we assumed that turbulence is isotropic such that the dimensionless vertical diffusion coefficient is just given by $\aturb$. 
In the code, another expression for $\hscdust$ is used which is essentially equivalent to the one in this equation (see Eq.~(51) of \cite{birnstiel2010} for details).
Since we use a one-dimensional model with $z$-integrated values, and the coagulation equation scales with the density squared, the Stokes number is calculated at mid-plane values where densities are the highest. Furthermore, we assume
we always stay in the Epstein drag regime \citep{epstein1924} which leads to \citep{birnstiel2010,Birnstiel2012}
\begin{equation}
	\label{eq:StEp}
	\stokes=\frac{\rhoint}{\rhog}\frac{\dustrad}{\vth}\Okepl\overset{z=0}{\approx}\frac{\pi}{2}\frac{\dustrad\rhoint}{\surfgas}.
\end{equation}
Since we use $\stmax=10$ throughout in this paper, one might think that in this case one would be in the Stokes drag regime \citep[see Eq.~(126) of][]{stokes1851}. The transition between Epstein and Stokes drag occurs if the gas mean free path is around $4/9$ the radius of 
the particle \citep{weidenschilling1977}. Since Stokes numbers of $10$ are only reached in the very outer part of the disk, where the gas density is sufficiently low,
the gas mean free path is many orders of magnitude larger than the size of any particle. Dependent on the gas density, dust grains don't have to grow by much, if at all, to reach these large Stokes numbers. The Stokes drag regime is only reached within a few $\AU$. But, again, we
will ignore this regime here to avoid overlapping effects with our model.

For growth and fragmentation we solve the Smoluchowski equation \citep{Smoluchowski1916}. It considers all binary combinations of particle size bins. It uses a probability for fragmentation and coagulation, depending on the collision speed. 
This method is identical to the one of \cite{birnstiel2010}, but more details on the Smoluchowski equation can be found in Appendix~\ref{sec:coag_app}.

The relative speed of collisions, $\vrelcoll$, can have different contributions. There are systematic velocities from radial drift \citep{weidenschilling1977,Nakagawa1986}, azimuthal drift \citep{Nakagawa1986}, and vertical settling \cite[p. 26]{Safronov1969}. 
These scale with the particles Stokes number and vanish for equally sized grains. On the other hand, random speeds such as 
turbulence \citep{OrmelCuzzi2007} and Brownian motion lead to non-vanishing relative velocities between equally sized particles. 
The relative turbulent velocities are proportional to $\sqrt{\aturb}$, thus $\aturb$ can be interpreted
as a measure not only of turbulence strength in terms of typical turbulent gas velocities, but also of particle collision speed. 

The radial gas velocity is given by \citep{lbp1974}
\begin{equation}
	\label{eq:ugasr}
	\vgask{r}=-\frac{3}{\surfgas\sqrt{r}}\pd{}{r}(\surfgas\nu\sqrt{r}).
\end{equation}

Since the gas motion is determined by a balance of forces among stellar gravity, the gas pressure gradient force, and the centrifugal force,
it orbits the star with a 
different speed than solids, which don't feel a pressure gradient force. Due to a lack in centrifugal force ($\stokes<1$), single 
particles drift radially with a velocity of \citep{adachi1976,weidenschilling1977}
\begin{equation}
	\vdrift=\frac{\stokes}{\stokes^2+1}\frac{\hscale}{r}\pd{\ln{\press}}{\ln{r}}\sound,
\end{equation}
with $\press=\sound^2\rhog$, {\ie}, assuming an isothermal case with adiabatic index $1$. 
Since radial gas motion drags dust along, the total radial dust velocity is given by \citep{TakeuchiLin2002}
\begin{equation}
 \label{eq:vdusttotal}
 \vdustk{r}=\frac{\vgask{r}}{1+\stokes^2}+\vdrift.
\end{equation}

We model the fragmentation probability as a smooth transition from $1$, for collision speeds above the breakup speed $\vfrag$, to $0$ for collision speeds below,
\begin{equation}
 \label{eq:fprob-smooth}
 \fprob:=
 \begin{cases}
  0.5\cdot\exp{\left[2(\vrelcoll-\vfrag)/(\mathrm{cm/s})\right]} &\vrelcoll\leq\vfrag\\
  1-0.5\cdot\exp{\left[-2(\vrelcoll-\vfrag)/(\mathrm{cm/s})\right]} &\vrelcoll>\vfrag
 \end{cases}
 ,
\end{equation}
where the radius dependent fragmentation velocity is modeled by the smooth function
\begin{equation}
 \label{eq:vfragm-r}
 \frac{\vfrag(r)}{\mathrm{m/s}}=
  \begin{cases}
   \displaystyle10^{\displaystyle0.5\cdot\exp{\left[4(r-\rice)/\mathrm{\AU}\right]}} & r\leq\rice\\
   \displaystyle10^{\displaystyle1-0.5\cdot\exp{\left[-4(r-\rice)/\mathrm{\AU}\right]}} & r>\rice
  \end{cases}
  .
\end{equation}
The coagulation probability is given via $\cprob+\fprob=1$. 
Experiments with silicate dust
grains measured velocities of $1\,\mathrm{m/s}$ for the onset of fragmentation \citep{BlumMuench1993, BlumWurm2008}, and theoretical studies \citep{leinhardt2009}
found similar values. It was found numerically \citep{wada2009} that the fragmentation velocity for $10\,\mathrm{\micron}$ sized
icy dust aggregates can reach $50\,\mathrm{m/s}$. Laboratory studies found threshold speeds about $10\,\mathrm{m/s}$
for icy grains \citep{GundlachBlum2014}. But this speed depends on the monomer size. Experiments show that $\vfrag$ decreases with grain size for silicate dust grains \citep{Beitz2011},
which could be partly attributed to the increasing influence of inhomogeneities with growing grain size.
For ice, measured erosion threshold velocities are around $15\,\mathrm{m/s}$ \citep{GundlachBlum2014}. However, we choose 
$\vfrag=10\,\mathrm{m/s}$ outside the water ice line. 
As already mentioned by \cite{birnstiel2010}, the decrease of the threshold collision speed within the ice line leads to a traffic jam effect since inward drifting particles are forced to fragment to smaller sizes, which in turn drift slower. 
%\begin{equation}
% \vbm(i,j)=\sqrt{\frac{8\kB T (m_i+m_j)}{\pi m_i m_j}}\;.
%\end{equation}
%\begin{equation}
% \Delta\vdustco_r(i,j)&=\frac{\sound^2}{\vkepl}\pd{\ln{\press}}{\ln{r}}\abs{\frac{1}{\stokes_i+\stokes_i^{-1}(1+\ratiodtog)^2}-\frac{1}{\stokes_j+\stokes_j^{-1}(1+\ratiodtog)^2}}\;,\\
% \Delta\vdustco_\varphi(i,j)&=\frac{1+\ratiodtog}{2}\frac{\sound^2}{\vkepl}\pd{\ln{\press}}{\ln{r}}\abs{\frac{1}{\stokes_i^2+(1+\ratiodtog)^2}-\frac{1}{\stokes_j^2+(1+\ratiodtog)^2}}\;,\\
% \Delta\vdustco_z(i,j)&=z\Okepl\abs{\stokes_i-\stokes_j}\;.
%\end{equation}
%\begin{equation}
%\label{eq:vrelturb-OrmelCuzzi2007}
% \Delta v_\mathrm{turb}(i,j)=\begin{cases}\sqrt{\aturb}\sound\sqrt{\dfrac{\stokes_i-\stokes_j}{\stokes_i+\stokes_j}}
%			 \sqrt{\dfrac{\stokes_i^2}{\stokes_i+\reynolds^{-1/2}} - \dfrac{\stokes_j^2}{\stokes_j+\reynolds^{-1/2}}}
%			 \quad &,\;\stokes_{i,j}\le\reynolds^{-1/2}%\ \forall\ \stokes_j<\stokes_i 
%			 \\ \\
%			 \sqrt{\aturb}\sound\sqrt{2.2\stokes_i-\stokes_j+\dfrac{2\stokes_i^2}{\stokes_i+\stokes_j}\left( \dfrac{1}{2.6}+\dfrac{\stokes_j^3}{1.6\stokes_i^3+\stokes_i^2\stokes_j} \right)}
%			 \quad &,\;\reynolds^{-1/2}\le\stokes_i\le1%\ \forall\ \stokes_j<\stokes_i
%			 \\ \\
%			 \sqrt{\aturb}\sound\sqrt{\dfrac{1}{1+\stokes_i} + \dfrac{1}{1+\stokes_j}}\quad &,\;1>\stokes_i%\ \forall j<i
%			 \end{cases}\;.
%\end{equation}

We follow \cite{YoudinLithwick2007} and model the radial particle diffusivity as
\begin{align}
	\diffd=\frac{\diffg}{1+\stokes^2}
\end{align}
and we assume that the radial gas diffusivity $\diffg$ equals the turbulent viscosity of Eq.~\eqref{eq:turbvisc}. 

Our initial but also fixed gas column density profile follows the self-similar profile as found by \cite{lbp1974},
\begin{equation}
 \label{eq:gasini}
 \surfgas(r)=\sigchar\left(\frac{r}{\rchar}\right)^{-\vispow}\exp{\left[-\left(\frac{r}{\rchar}\right)^{2-\vispow}\right]}.
\end{equation}
% Let f(x,t) be a function, f:x\in R,t\in R-> R and alpha, beta, xi \in R. A solution to a differential equation is called self-similar if 
% f(x,t)=t^alpha*eta(xi), xi=xt^beta
% where eta(xi) \in R is a real function. I.e., the solution from Lynden-Bell and Pringle is indeed self-similar. We just fixed the time
% parameter.
The normalization constant is given by 
\begin{align}
 \sigchar=(2-\vispow)\Mdisk/[2\pi\rchar^2\cdot(1+\dtog(t_0))]. 
\end{align}
At the so-called characteristic radius $\rchar$, the transition between the power-law and the exponential law occurs. 
The initial profile of the total dust amount is assumed to follow 
\begin{align}
 \dustsurf(t_0)=\dtog(t_0)\surfgas(t_0).
\end{align}
The initial dust-to-gas ratio
is set to $\dtog(t_0)=0.01$ for all simulations, as found for the interstellar 
medium \citep[\eg,][]{savage1972}. 
We also fix $\vispow=1$ and $\rchar=35\,\mathrm{AU}$ in this paper. Furthermore, 
we keep the static gas column density profile and do not let 
the disk (viscously) evolve. This has the advantage that we can see the effects of our model parameters for 
planetesimal formation more clearly without overlapping effects from gas mass transport. Even though we don't let $\surfgas$ evolve, we allow a radial gas velocity according to 
Eq.~\eqref{eq:ugasr}. This leads to outward transport of small grains in the outer disk. For our model parameters, this occurs roughly from $20\,\AU$ to $200\,\AU$ and below a 
particle Stokes number of ($\stokes\ll1$)
\begin{align}
 \begin{aligned}
 \label{eq:stgasdom}
 \stokes&\leq
 \dfrac{6\abs{2-\expT-\vispow-(2-\vispow)\left(\dfrac{r}{\rchar}\right)^{2-\vispow}}}
 {2\vispow+3+\expT+2(2-\vispow)\left(\dfrac{r}{\rchar}\right)^{2-\vispow}}\aturb\\
 &\overset{\vispow=1}{\underset{\expT=0.4}{=}}\dfrac{6\abs{0.6-r/\rchar}}
 {5.4+2r/\rchar}\aturb
 \end{aligned}
\end{align}
as derived in Appendix~\ref{parvel}. 
% The initial dust column density is then $\surfgas$ times the initial dust-to-gas ratio which is set to $\dtog=0.01$ for all simulations as found for the interstellar 
% medium \citep[\eg,][]{savage1972}. 
For the initial size distribution we assume the dust to follow the distribution in the interstellar medium found by \cite{mathis1977}. \Ie, the number density $\numdens$ is described by the power-law
\begin{equation}
	\label{eq:nini}
	\numdens(\dustrad)\propto\dustrad^{-3.5}
\end{equation}
with dust radii ranging from $\dustrad=0.1\,\mathrm{\micron}$ to $1\,\mathrm{\micron}$. 
For a review on dust evolution in circumstellar disks see, {\eg}, \cite{birnstiel2016}.

We fix the temperature to a profile similar to that of a radiative equilibrium disk \citep[they derived $T\propto r^{-3/7}$]{chiang1997} 
\begin{equation}
	\label{eq:Tprofile}
	T(r)=158.9\,\mathrm{K}\cdot\left(\frac{r}{\AU}\right)^{-2/5}. %63.25\,\mathrm{K}\cdot\left(\frac{r}{10\,\AU}\right)^{-2/5}
\end{equation}
The ice line is defined as the position where $T=170\,\mathrm{K}$ in this 
paper. Within the ice line we reduce the mass of particles to $1/3$ 
of their value for planetesimal formation, {\ie} we assume an ice mass fraction 
of $2/3$. For this, we define the 
rocky-to-(ice and rocky) mass parameter 
\begin{align}
	\label{eq:dice}
	\dice = 1/3\cdot\theta(T-170\,\mathrm{K}) + \theta(170\,\mathrm{K}-T).
\end{align}
\cite{lodders2003} summarizes her findings in Table 11 where the total rock/(total rock+H$_2$O ice) 
ratio gives $0.46$. In section 3.3 of \cite{min2011} they find values around $0.54$. 
\cite{Hayashi1981} uses a value of $0.24$ for this ratio. 
We choose a value in between by assuming $1/3$ but in the results it is only important to notice the kink in the final planetesimal distribution at the water ice line. 
For the particle dynamics and growth processes this reduction in mass 
is not considered. Within the water ice line our simulation thus has 
too much mass in particles, but the maximum particle size is only weakly 
affected by this since, in the fragmentation limit, particles grow
to the limit again faster than they drift inward. 
To be able to analyze the impact of our planetesimal model parameters, we do not include accretion heating \citep{Pringle1981}, opacity effects, or radiative evolution.

\subsection{Parameterized Planetesimal Formation}
Do we need to restrict the particle species contributing to planetesimal 
formation? This is the question we are facing in this section. 
Since particle dynamics
depend on the Stokes number, one could think 
that there exists an interval $[\stmin,\stmax]$ between which particles 
have the right properties for trapping and building clumps of Hill density. 
The pebble flux
regulates the rate at which material is delivered, so this does not need to be restricted by a Stokes number interval.
The question is whether there exist particles which cannot participate in trapping \emph{and} planetesimal formation. 
We know that both trapping,
either in zonal flows \citep{Dittrich2013} or vortices \citep{fu2014,Raettig2015}, and collapse to planetesimals \citep{andydiss} work 
for particles with $\stokes=10^{-2}$. 
% But we don't know whether this still works for smaller particles. 
From \cite{Carrera2015} we learned that $\stmax\approx10$. \cite{yang2017} have shown that streaming instability is possible for $\stokes=10^{-3}$ particles if $\dtog\geq4\cdot10^{-2}$. 
We know that particles with $\stokes=10^{-2}$ still contribute to trapping and gravitational collapse to planetesimals. But so far, we cannot entirely exclude that smaller particles 
can take part in this process as well---especially for mixtures of 
different particle species with some mass distribution depending on their 
Stokes number. However, we use $\stmin=10^{-2}$ and 
$\stmax=10$ in this paper.

The velocity field of a vortex or zonal flow
is modified 
around pressure bumps, and thus gas motion here differs from the gas motion in a smooth 
pressure structure. We assume that this motion can be neglected in 
comparison to the drift velocity. 
%  To distinguish: collapse of just $\stokes<10^{-2}$ particles and a mixture of 
%   particles including larger ones. Compare their collapse with diffusion and shear time scale.

We further assume that these structures have an average lifetime $\tlife$ 
proportional to the orbit time. It has been shown that both zonal flows 
\citep{Dittrich2013} and vortices \citep{MangerKlahr2018} can endure up to 
hundreds of orbits. Hence, we assume $\tlife=100\,\torb$ in this paper. 
Based on work by \cite{Dittrich2013} we assume an average radial separation of $\trapdist=5\hscale$.

\begin{figure*}[th]
  \centering
  \includegraphics[width=.9\textwidth]{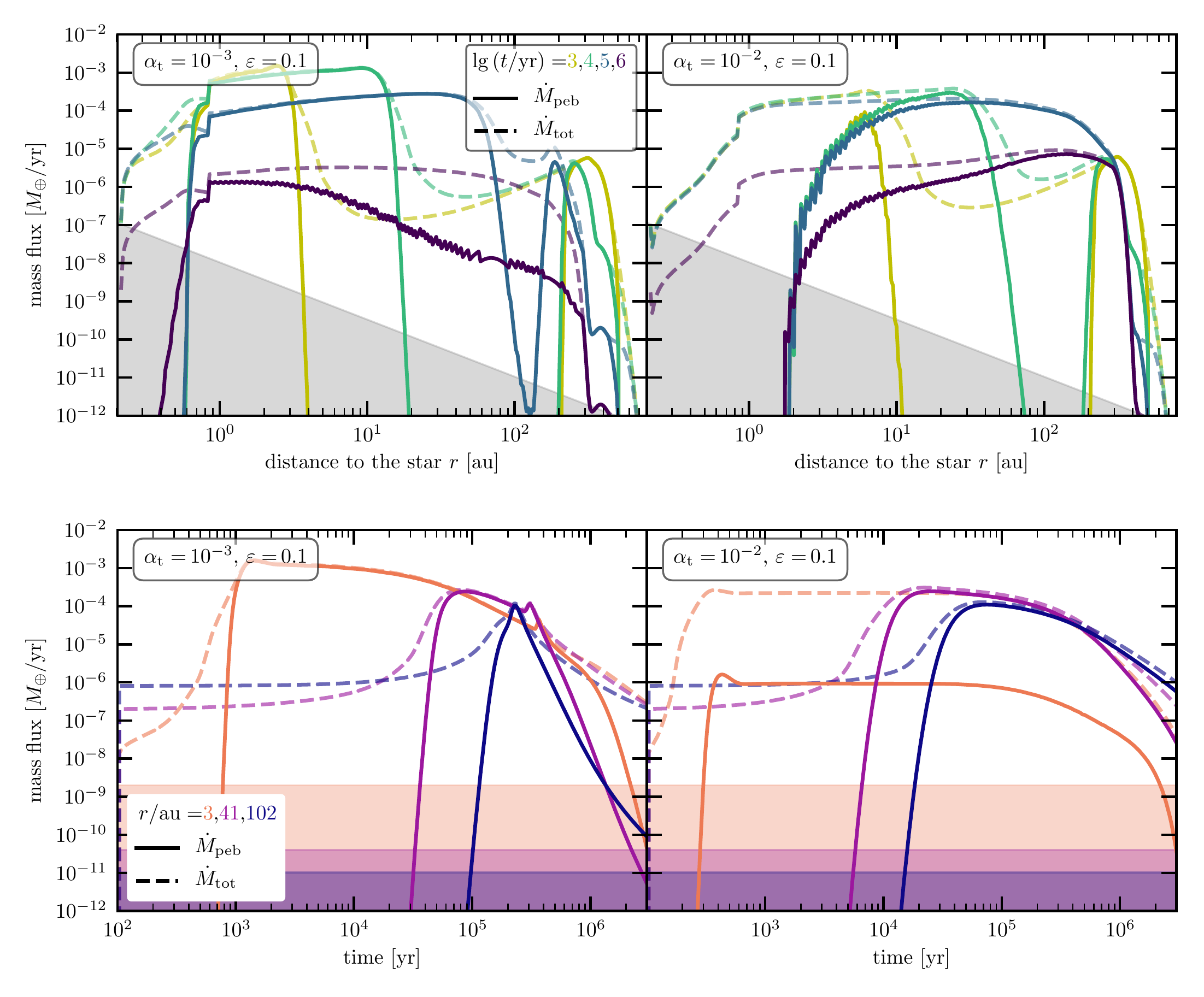}
  \caption
  {Local pebble flux $\pebflux$ of particles in the range $10^{-2}<\stokes<10$ in Earth masses per year (solid lines) for two different assumptions of the disk turbulence: $\aturb=10^{-3}$ (left panels) and $\aturb=10^{-2}$ (right panels), both at a trap efficiency $\peff=0.1$. 
  The upper panels present a time series of the pebble flux as a function of disk radius for times from $10^3$ (yellow) to $10^6\,\mathrm{years}$ (purple) in steps of decadic factors. 
  The lower panels show the evolution at different radii as a function of time at $3$ (orange), $41$, and $102\,\AU$ (dark blue). 
  In addition we overplot the total mass flux $\totflux$, i.e. pebbles plus smaller grains (dashed lines).  In the lower panels, within the shaded areas and below, the mass flux is smaller than the critical particle flux $\Mcr$ below which planetesimal formation does not occur in this model. This is marked by the gray zone in the upper panels.
   }
  \label{fig:pebflux}
\end{figure*}
Our analytical description of planetesimal formation arises in the code 
as a sink term in the 
advection-diffusion equation for the column density of a single particle species of mass $m_i$, given by
\begin{align}
\begin{aligned}
	&\pd{\dustsurf^i}{t}+\frac{1}{r}\pd{}{r}
    \left\{ 
    	r\left[  
        		\dustsurf^i\vdustk{r}^i-\diffd^i\pd{}{r}\left(
                										\frac{\dustsurf^i}{\surfgas}
                									  \right)
                \surfgas
        	  \right]
    \right\}
    \\
    &=-\frac{\peff}{\trapdist}\abs{\vdrift^i}\dustsurf^i
       \cdot\theta(\pebflux-\Mcr)
    \\
    &\hphantom{=\;\;}\theta(\stokes_i-\stmin)\cdot\theta(\stmax-\stokes_i)
    .
\end{aligned}
\end{align}
The Heaviside functions $\theta(\cdot)$ represent our conditions that the pebble flux must be critical and that only particles with 
$\stmin\leq\stokes\leq\stmax$ are allowed to build planetesimals. 
We use the flux conserving implicit donor-cell scheme from Appendix~A.1 of \cite{birnstiel2010}, where we set
\begin{align}
\begin{aligned}
    L_i&=-\frac{\peff}{\trapdist}\abs{\vdrift^i}\cdot\theta(\pebflux-\Mcr)\\
       &\hphantom{=\;\;}                    \theta(\stokes_i-\stmin)\theta(\stmax-\stokes_i)
\end{aligned}
\end{align}
in their Eq.~(A.1) in order to have an implicit loss term due to planetesimal formation. 
The number of new planetesimals created within the time step $\dif{t}$
is then given, in terms of column density, by
\begin{align}
	\dif{\surfp}=\dotsurfp\dif{t}=-\sum_{\stmin\leq\stokes_i\leq\stmax} L^i\dif{t}.
\end{align}
The pebble flux $\pebflux$ is estimated from the last time step, but the resulting error from this is very small.
% There must also be a minimum (Stmin) and maximum (Stmax) Stokes number of the particles,
% which are able to get trapped in vortices or zonal flows and participate in the gravitational collapse,
% since they must be large enough to sufficiently decouple from the gas and small enough such
% that they can be trapped without leaving the traps before they are able to participate in this
% planetesimal formation scenario.
% The contributing
% particles in zonal flows and vortices are roughly in a range of $\stmin= 10^{-2}$ to $\stmax = 10$ \citep{Barge1995, Dittrich2013} and
% the lifetimes of zonal flows are around
% $100\torb$ \citep{Dittrich2013} and $\torb = 2\pi/\Okepl$ is the orbital period.
% \section{Test cases}
% As test cases we show the cases of strong turbulence ($\aturb=10^{-2}$) and moderate turbulence ($\aturb=10^{-3}$) combined with 
% high ($\peff=0.8$) and low efficiencies ($\peff=0.1$).
% We have tested our model in simulations with $\peff=0.1$ and $0.8$ (feeding zone \& slope changed, depends on size distribution) and also varied the turbulence parameter 
% $\aturb=10^{-2}$ and $10^{-3}$ (size distribution changed).
\\ \\ \\
\section{Results}\label{results}
\begin{figure}[th]
  \centering
  \includegraphics[width=.48\textwidth]{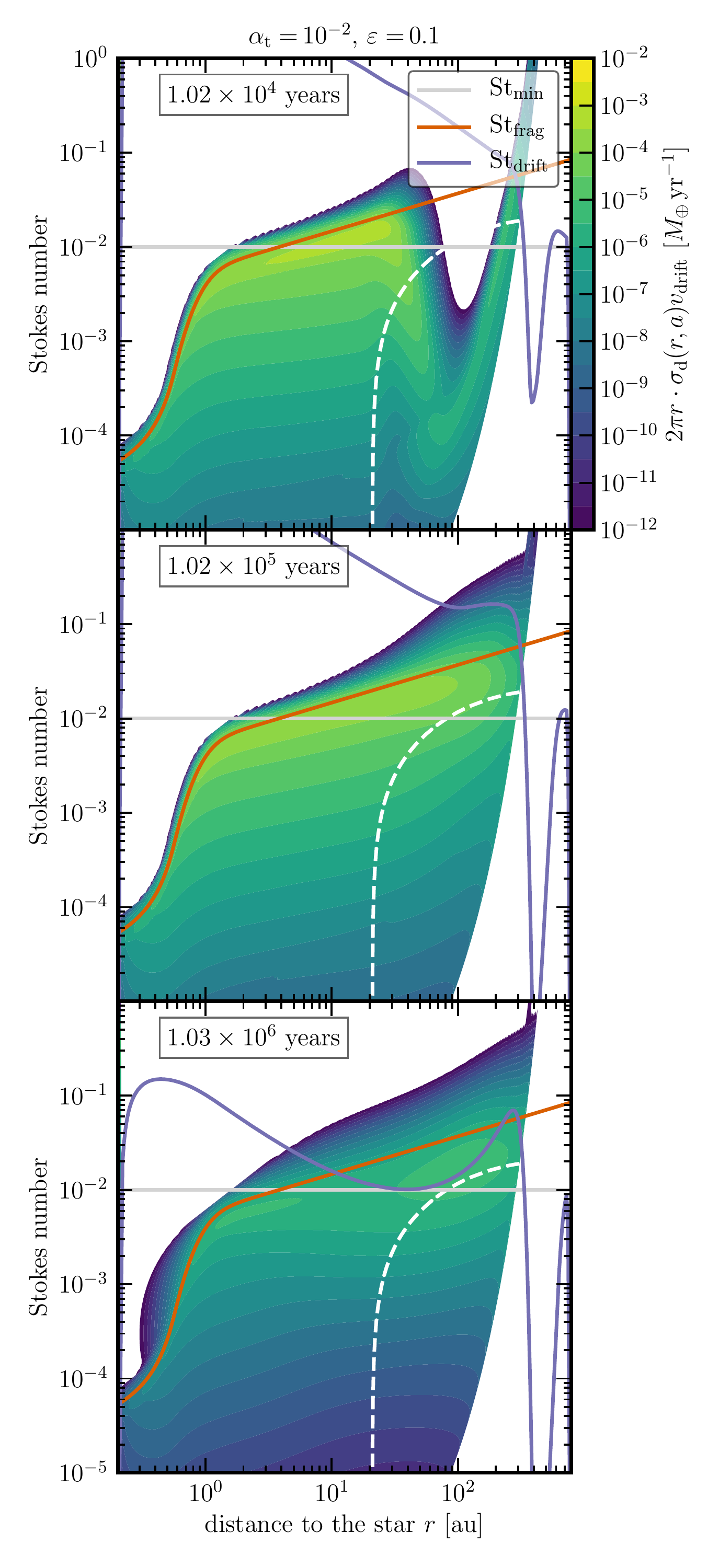}
  \caption
  {
  Local particle flux due to the radial drift velocity per size bin (color) as a function of Stokes number and disk radius. The turbulence level is $\aturb=0.01$ with an efficiency parameter for planetesimal formation of $\peff=0.1$. 
%   For high $\aturb$ (here $10^{-2}$), growth is limited by fragmentation events and leads to a less top-heavy size 
%   distribution compared to the drift limited case. Hence, the replenishment of small, slow drifting particles leads to a 
%   longer timespan of planetesimal formation well ordered by $\peff$.
  Particles below the white dashed line move outward. Even though we only show the
  flux resulting from $\vdrift$, %the radial drift velocity, 
  for the white dashed 
  line we also considered the gas velocity according to 
  Eq.~\eqref{eq:vdusttotal}. For this line, motion due to diffusion is not considered but it is included in the simulation. The horizontal gray line shows the lower limit 
  for which we allowed the particles to participate in the gravoturbulent 
  planetesimal formation process. The purple and red lines show the drift (Eq.~\eqref{eq:stdrift}) and fragmentation barrier (Eq.~\eqref{eq:stfrag}), respectively. 
  In Appendix~\ref{origin_part_grt_growth_barriers} we explain where the 
  particles greater than the growth barriers stem from.
  }
  \label{fig:s01_e01_r35_M5_flux_St_4_5_6_nodiskev}
\end{figure}
\begin{figure}[th]
  \centering
  \includegraphics[width=0.48\textwidth]{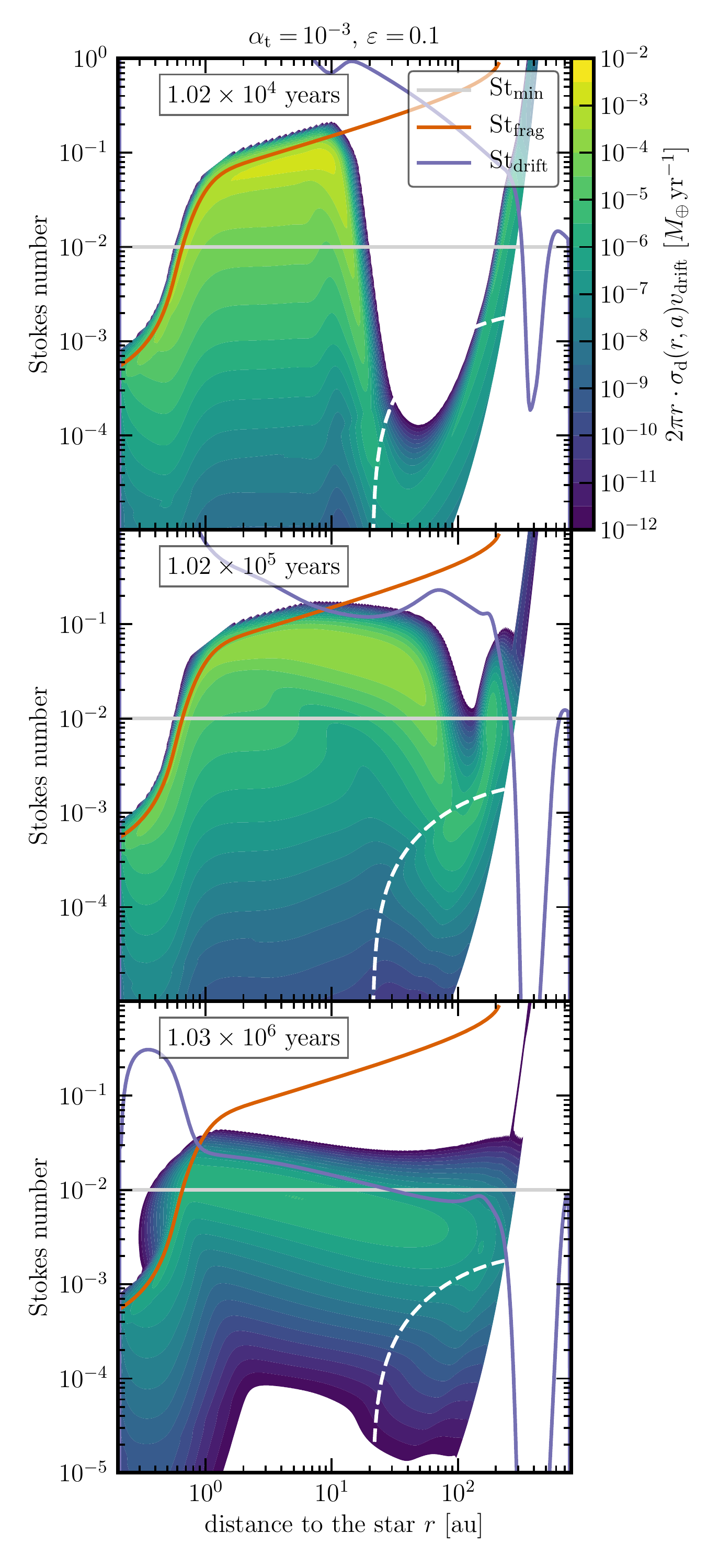}
  \caption
  {
  Same as Fig.~\ref{fig:s01_e01_r35_M5_flux_St_4_5_6_nodiskev} but with $\aturb=10^{-3}$ and $\peff=0.1$. Since the 
  fragmentation barrier is inversely proportional to $\aturb$, the maximum size and thus also the maximum Stokes number is 
  much larger. Most of the time the disk is primarily limited by drift which causes a top-heavy size distribution. Growth is slower compared to the $\aturb=0.01$ simulation because of smaller relative turbulent velocities leading to lower collision rates.
  }
  \label{fig:s02_e01_r35_M5_flux_St_4_5_6_nodiskev}
\end{figure}
\begin{figure*}[th]
  \centering
  \includegraphics[width=.8\textwidth]{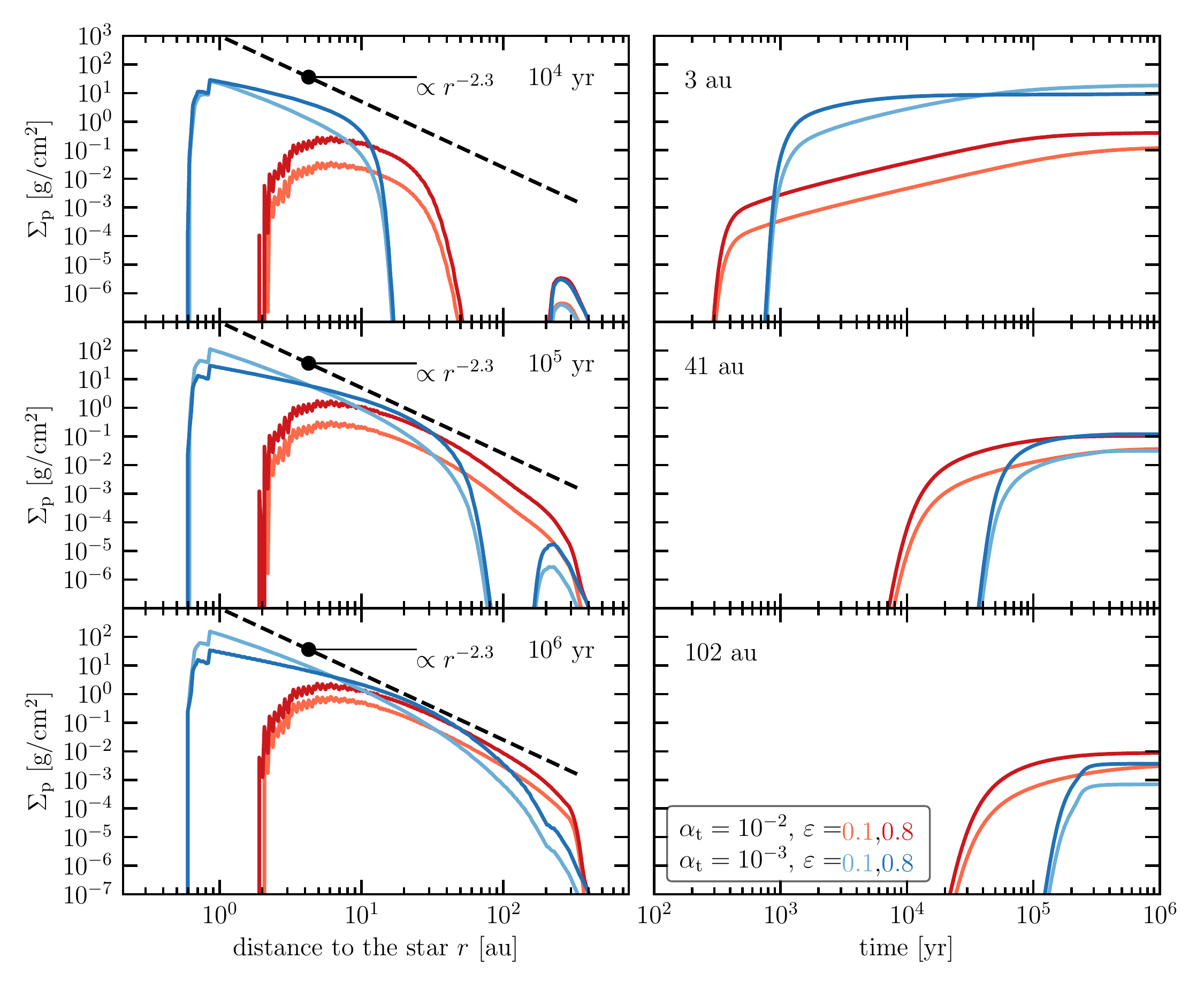}
  \caption
  {
   Planetesimal column density for different snapshots (left panels) vs. disk radius and for different radial disk positions as a function of time (right panels). 
   The fact that the final $\surfp$ at $40\,\AU$ (see middle right panel) of 
   the $\aturb=10^{-2}$ and $10^{-3}$ simulations is roughly the same for the same $\peff$ is pure coincidence. The initial 
   characteristic radius was 
   $\rchar=35\,\AU$ and the initial total disk mass $\Mdisk=0.05\,\msun$. 
   The black dashed line shows the predicted slope for a r-independent 
   pebble flux after Eq.~\eqref{eq:ptesprop}.
   The zigzag structure which can be seen within $10\,\AU$ for $\aturb=0.01$ stems from a lack 
   of mass resolution. This
   is also visible in Fig.~\ref{fig:s01_e01_r35_M5_flux_St_4_5_6_nodiskev} above the $\stmin$ line.
   The sharp kink at large radii for $\aturb=10^{-2}$ is caused by $\stokes(0.1\,\mathrm{\micron})=\stmax$, \ie, by a rapidly 
   decreasing pebble flux. A viscously evolving disk with dispersal would smear out this kink. 
   }
  \label{fig:ptes_nodiskev}
\end{figure*}

\begin{figure*}[th]
  \centering
  \includegraphics[width=.8\textwidth]{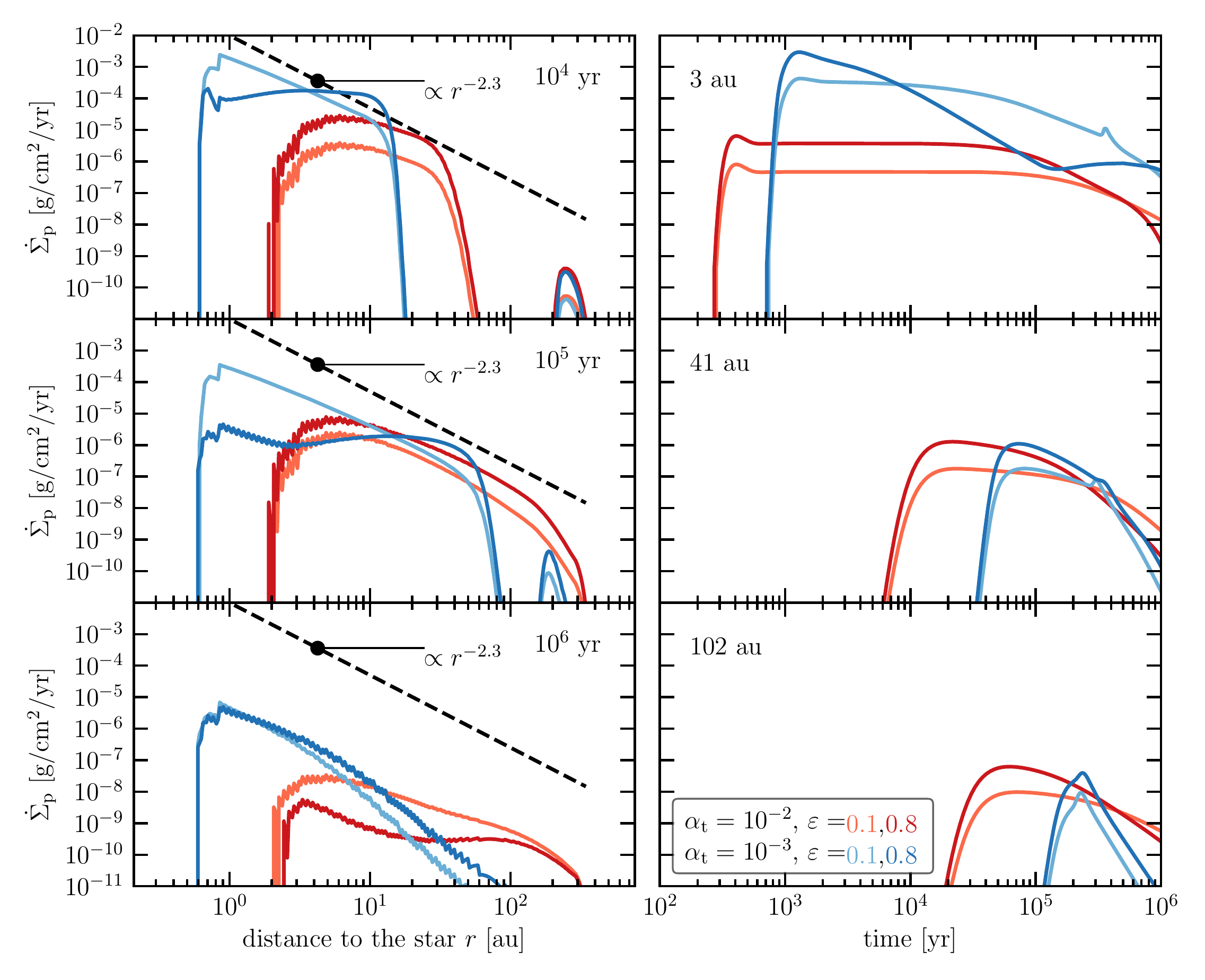}
  \caption
  {
   Same as Fig.~\ref{fig:ptes_nodiskev} but showing the planetesimal column density formation rate $\dotsurfp$.
   }
  \label{fig:ptesrate_nodiskev}
\end{figure*}

\begin{figure*}[th]
  \centering
  \includegraphics[width=.8\textwidth]{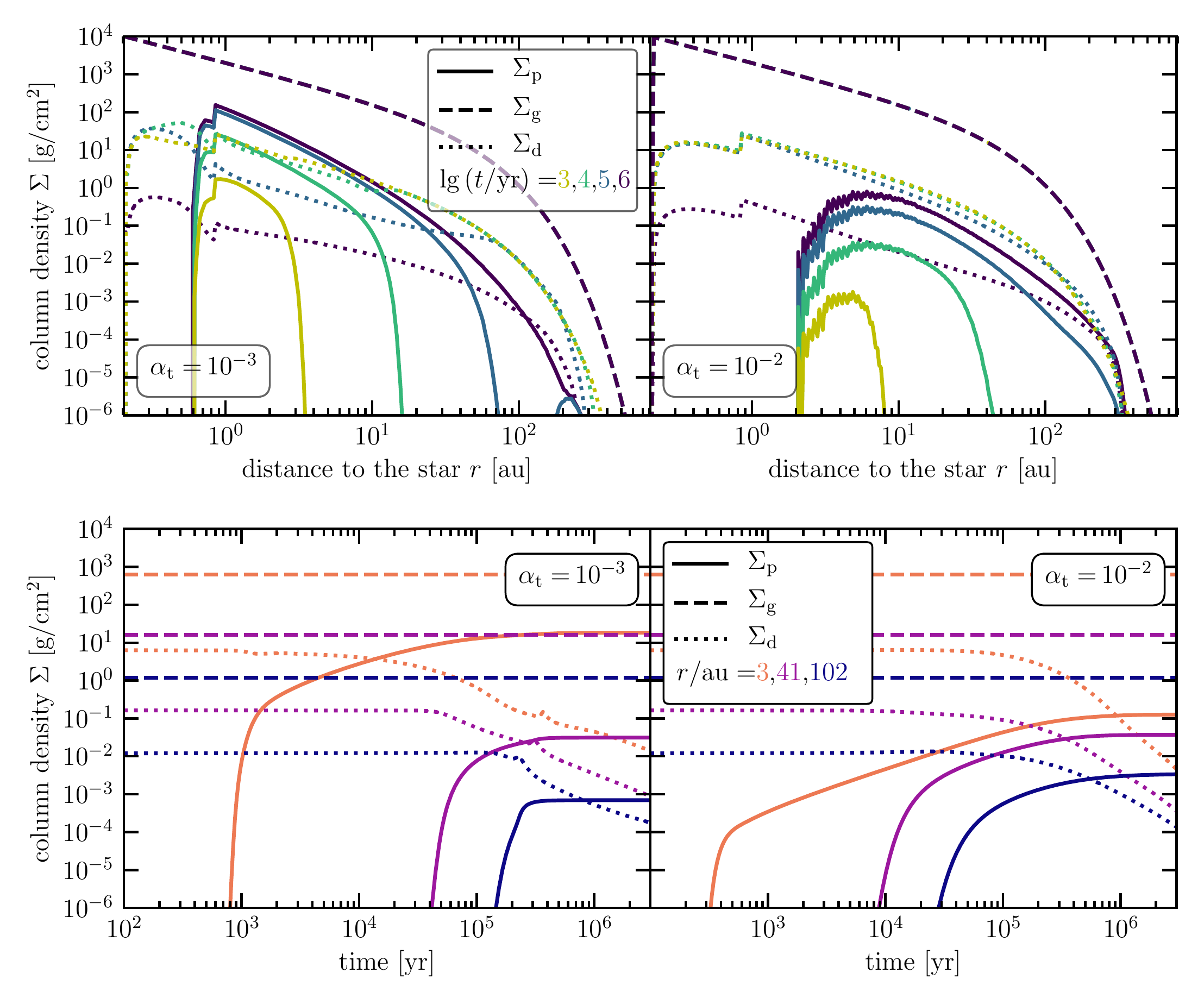}
  \caption
  {
  Evolution of the gas (dashed lines), dust (dotted lines), and planetesimal (solid lines) column densities for 
  simulations with $\aturb=10^{-3}$ (left panels) and $\aturb=10^{-2}$ 
  (right panels) where $\peff=0.1$. The upper panels show a time series of the column densities as 
  a function of radius and the lower ones the densities at local positions 
  as a function of time. 
%   The gray lines in the upper panels denote the minimum mass Solar Nebula. 
  The characteristic kink around 
  $0.8\,\mathrm{AU}$ is caused by the ice line, which does not move because the temperature $T(r)$ is not allowed to change. 
%   The legends in the upper and lower panels are valid for both panels in the same row.
  }
  \label{fig:sigma_a-2a-3_nodiskev}
\end{figure*}
We start our analysis with the pebble flux because it is directly linked to our planetesimal formation. 
Particle dynamics is mostly determined by the growth barriers \citep{Birnstiel2012}, {\ie} the fragmentation barrier
\begin{align}
    \label{eq:stfrag}
	\stfrag=\frac{1}{3\aturb}\left(\frac{\vfrag}{\sound}\right)^2
\end{align}
and the drift limit (here in the Epstein regime only)
\begin{align}
    \label{eq:stdrift}
	\stdrift=\frac{\dustsurf}{\surfgas}\left(\frac{\vkepl}{\sound}\right)^2\abs{\dfracpd{\ln{\press}}{\ln{r}}}^{-1}.
\end{align}
As can be seen in 
Fig.~\ref{fig:pebflux}, for stronger turbulence levels (here 
$\aturb=10^{-2}$) the particle flux stays high for longer times because 
particles are smaller and thus drift slower, see Fig.~\ref{fig:s01_e01_r35_M5_flux_St_4_5_6_nodiskev}. 
Here we defined the dust column density distribution per logarithmic bin of grain radius $\dustrad$ as
\begin{align}
	\sigma_\mathrm{d}:=\int_{-\infty}^\infty \nsize(\dustrad,r,z)\dustmass(\dustrad)\dustrad\dif{z}
\end{align}
with $\nsize$ being the number density per grain size bin. The total dust column density can then be written as
\begin{align}
	\dustsurf(r)=\int_{-\infty}^\infty\sigma_\mathrm{d}(r,\dustrad)\dif{\ln\dustrad}.
\end{align}
In the early stages, 
the flux is smaller for lower $\aturb$
for the same reason. The pebble 
growth front, which is the radius at which $\stmin$ particles form the first time, moves from the inside to the outside, while particles
drift inward from the outer disk because their Stokes number is 
high. When these two fronts clash, the density suddenly rises and a 
particle wave propagates inward which is why one can see a small local peak 
in the curves of the lower left panel. In the lower right this is not 
visible since fragmentation is damping this effect. 

At late times ($\sim10^6\,\yr$) the pebble to total mass flux ratio drops 
significantly in the $\aturb=10^{-3}$ case. This is due to the drift barrier, which 
drops to smaller sized particles since $\dustsurf$ is decreasing, see 
Fig.~\ref{fig:s02_e01_r35_M5_flux_St_4_5_6_nodiskev}. The gray zone in the 
upper panels mark the sub-critical flux values, {\ie}, the fluxes that are too low to allow formation of planetesimals within a trap lifetime. One could change $\tlife$ or $\mptes$ 
by a factor of $10$ without changing the results significantly because 
the flux is orders of magnitude larger then the critical value.

   Since the mass reservoir is in the outer disk where gas densities are very low, the pebble flux 
   in the outer region around $\sim200-400\,\AU$ already reaches critical values, and leads to planetesimal formation, in the early phase ($\lesssim10^4\,\yr$). When the slope of $\surfp$ is steeper than $r^{-2}$ (see Fig.~\ref{fig:ptes_nodiskev}), the mass contribution to the total planetesimal 
   population in the disk will not 
   rise with $r$. \Ie, the contribution becomes less the further out
   in the disk the considered regions are. For a fixed 
   logarithmic binning, this can be seen
   by writing the planetesimal mass between disk radius $r_1$ and $r_2$ as
   \begin{align}
   		M_\mathrm{p}(r_1,r_2)=2\pi\int_{\ln{r_1}}^{\ln{r_2}}\surfp r^2\dif{\ln{r}}.
   \end{align}
   Hence, the probably unphysical outer bump in $\surfp$ doesn't contribute significantly if the slope is steeper than 
   $r^{-2}$. We think this outer formation region would not exist in the \emph{early} evolution of the disk, since 
   the formation time of structures such as vortices means they need a few orbits to form, which would take much longer in the outer than in the inner disk.
   
   If turbulence is strong (here $\aturb=10^{-2}$), fragmentation will be the growth limiting process.
   In that case not only is the maximum grain size
   smaller, but there are also more small grains in 
   general. Hence, in 
   comparison to simulations with weaker turbulence (here $\aturb=10^{-3}$), particles 
   in the outer disk stay there for longer and can form planetesimals. 
   Fig.~\ref{fig:ptes_nodiskev} and \ref{fig:ptesrate_nodiskev} show the 
   resulting planetesimal column density $\surfp$ and formation rate 
   $\dotsurfp$, respectively, both for different snapshots showing the 
   entire disk and for local evolutions. For the latter, we chose $3\,\AU$ 
   ($\sim$ position of the main Asteroid Belt), $40\,\AU$ ($\sim$
   position of the Kuiper Belt), and $100\,\AU$ to show the behavior in the 
   outer disk. Planetesimal formation basically starts shortly after the 
   first pebbles have formed. We summarize the effects as follows:
   \begin{itemize}
   		\item $\aturb=10^{-2}$: higher $\peff$ generally lead to higher planetesimal formation rates but these also 
        decline faster. 
   In the end this nevertheless ends in a more abundant planetesimal population for higher $\peff$ in the entire disk. 
   		\item $\aturb=10^{-3}$: in the outer disk $\dotsurfp$ is also higher for larger $\peff$, such as in the higher $\aturb$ case. 
   Radial drift removes the largest particles of the top-heavy size distribution. The
   fewer particles were converted into 
   planetesimals in the outer disk, the more material is available at later times in the inner regions (conservation of mass). 
   This can lead to more planetesimals in the inner disk for smaller $\peff$. 
   This effect cannot be seen in the results of the $\aturb=10^{-2}$ simulations since fragmentation events slow down the 
   radial motion and the feeding zones of planetesimals are much narrower. 
   \end{itemize}
   
   If the mass flux of pebbles is roughly 
   $r$-independent, one can estimate the slope of the outcoming planetesimal profile as we show in section~\ref{sec:slope}. 
   
   In Fig.~\ref{fig:sigma_a-2a-3_nodiskev} we show the column densities of gas, dust and planetesimals for
   different 
   snapshots and locations in the disk. This way one can not only see that for $\aturb=10^{-3}$ plantesimals form at 
   $2-3\,\AU$ after a few 
   thousand years,
   but also after a few ten to hundred thousand years at $20-30\,\AU$. Fig.~\ref{fig:sigma_a-2a-3_nodiskev} also shows the ratios
   of dust to gas, 
   planetesimals to dust and planetesimals to initial dust. For $\aturb=10^{-2}$, the zone where planetesimals are formed 
   moves faster outward due to faster growth from dust to pebbles. 
   We summarize the different ratios as follows: 
   \subsubsection*{\lowercase{dust-to-gas ratio} $\dustsurf/\surfgas$}
   This ratio is constantly decreasing since we kept the gas density constant. 
   In the early stages of evolution ($\lesssim10^6\,\yr$) $\dustsurf/\surfgas$ decreases faster for smaller $\aturb$ because particles drift faster.
%   \begin{itemize}
   	% 	\item dust-to-gas $\dustsurf/\surfgas$: this ratio is constantly decreasing since we kept the gas density constant. 
    %     In the early stages of evolution ($\lesssim10^6\,\yr$) $\dustsurf/\surfgas$ decreases faster for smaller $\aturb$ 
    %     because particles drift faster. 
        % \item planetesimals-to-dust $\surfp/\dustsurf$: 
        \subsubsection*{\lowercase{planetesimals-to-dust ratio} $\surfp/\dustsurf$}
        \begin{itemize}
        \item $\aturb=10^{-3}$: from $10^4\,\yr$ on, values above unity are 
        reached within $2\,\AU$. At $10^6\,\yr$ this ratio is $>1$ almost 
        everywhere in the disk. 
        \item $\aturb=10^{-2}$: only at late times ($\gtrsim10^6\,\yr$) values 
        above unity are reached.
        \end{itemize}
        % \item planetesimals-to-initial dust $\surfp/\dustsurf(t_0)$:
        \subsubsection*{\lowercase{planetesimals-to-initial dust ratio $\surfp/\dustsurf(t_0)$}}
        \begin{itemize}
        \item $\aturb=10^{-3}$: at $10^5-10^6\,\yr$ within $10\,\AU$ 
        values above unity are reached, indicating a larger feeding zone of 
        planetesimals, {\ie} material significantly contributing to 
        planetesimals can come from far outside. 
        \item $\aturb=10^{-2}$: the ratio never becomes greater than $1$ which 
        indicates that fragmentation allows more time to 
        form planetesimals out of the available material, {\ie} the drift 
        timescale of single particles \emph{including} 
        destruction events is slower than the timescale of 
        planetesimal formation.
        \end{itemize}
\subsection{Necessity for lower turbulence ($\aturb$) to form Earth}
If our assumptions are correct, no planetesimals are formed within 
$2\,\AU$. The reasons for that are on the one hand the value of $\stmin$ 
and on the other hand the temperature profile. The latter dictates the 
radial shape of the fragmentation barrier. For a more realistic temperature 
model, accretion heating would lead to higher temperatures in the inner disk 
and a steeper slope. As a result of that the water ice line would be 
shifted further out. Hotter temperatures lead to a larger gas scale height 
and thus also to larger $\trapdist$. At the same time particles would be 
smaller due to the lower fragmentation barrier. Both effects would cause 
a planetesimal free zone, or at least a much smaller planetesimal 
population in the inner disk ($\lesssim3\,\AU$). {\Ie}, these problems we see 
here would be even worse, which makes 
it very likely that our conclusion still holds if more physics is taken 
into account. As we show in Appendix~\ref{sec:totdiskmass}, the total 
mass of planetesimals in the entire disk is also problematic if 
$\aturb$ is high.

\cite{yang2018} measured in shearing box simulations with non-ideal MHD and a dead zone values of the order of $\aturb\sim10^{-4}$ to $10^{-3}$. In layered accretion theory values of $\aturb\sim10^{-6}$ to $10^{-4}$ seem to be preferred \citep{Johansen2015}. Hence, our finding is congruent with their findings.
% As long as the particle flux is \emph{not} dominated by particles in the range $[\stmin,\stmax]$ the result is very sensitive to the choice of these parameters. 

\subsection{Planetesimal column density steeper than that of initial gas and dust}\label{sec:slope}
Fig.~\ref{fig:sigma_a-2a-3_nodiskev} shows that for lower turbulence (here 
$\aturb=10^{-3}$) the slope of the planetesimal column density can be 
steeper than that of initial gas and dust. 
This result stems mostly from the $r$-dependency of $\trapdist\propto\hscale$. On top of this, transport phenomena can change it as well. For $\peff=0.1$, this can
even be shown analytically since the 
pebble flux does not depend on $r$ over a larger region of the inner disk 
(Fig.~\ref{fig:pebflux}). 
We find that the final slope of the planetesimal column density %in the inner disk ($\lesssim??\,\AU$) 
is steeper than
that of the gas if the effective efficiency $\peff/\trapdist$ is 
sufficiently small (here $0.02/\hscale$, $\peff=0.1$). If this effective efficiency is large (\eg, $0.16/\hscale$, $\peff=0.8$), the feeding zone of planetesimals is much smaller, \ie, 
planetesimal material stems from a local region, leading to a slope much closer to the initial dust density, which is the same as for the gas in our simulations. 
% \section{Analytical planetesimal column density profiles}

If the pebble flux is constant for all $r$, \ie, for size distributions dominated by particles with Stokes numbers $\stmin\leq\stokes\leq\stmax$ and sufficiently small 
$\peff/\trapdist$, then we can conclude that 
\begin{align}
 \label{eq:surfppropto}
 \surfp\propto\dotsurfp\propto\frac{v_1\surfpeb}{\hscale}\,
\end{align}
where $v_1$ is the pebble drift speed. 
% \begin{align}
%  v_1\propto\frac{\surfpeb}{\surfgas}r^{-1/2}
% \end{align} 
% is the drift velocity of the representative particles dominating the flux according to \cite{Birnstiel2012}. 
For a constant pebble flux, $\pebflux$, the pebble column density is given by
\begin{align}
 \label{eq:surfpebpropto}
 \surfpeb=\frac{\pebflux}{2\pi}\frac{1}{r v_1}
\end{align}
% which gives in the drift limit a pebble column density profile of
% \begin{align}
%  \Sigma_\mathrm{peb,drift}\propto\sqrt{\frac{\surfgas}{r^2\Okepl}}
% \end{align}
% as both found by \cite{Birnstiel2012}. We obtain
% \begin{align}
%  \label{eq:ptesprop}
%  \surfp\propto\frac{{\Sigma_\mathrm{peb,drift}}^2}{\surfgas}\frac{\Okepl}{\sqrt{r}\sound}
%  \propto r^{(\expT-5)/2}.
% \end{align}
leading to
\begin{align}
 \label{eq:ptesprop}
 \surfp\propto\dotsurfp\propto1/(r\hscale)
 \propto r^{(\expT-5)/2}.
\end{align}
% This result can also be directly obtained from Eqns.~\eqref{eq:surfppropto} and \eqref{eq:surfpebpropto}, $\dotsurfp\propto1/(r\hscale)$. 
For the temperature power-law index we used throughout this paper, $\expT=0.4$, this gives a planetesimal column density profile of 
$\surfp\propto r^{-2.3}$,
independent of the gas column density profile. In this case, the {\em slope} of the planetesimal column density $\surfp$ 
does not depend on $\peff$, but the total value of $\surfp$ scales with it linearly. 
For regions limited by fragmentation, the condition that the flux must be dominated by pebbles ($\stmin\leq\stokes\leq\stmax$) is almost always violated 
since fragmentation grinds particles to sizes with Stokes numbers smaller than $\stmin$. Due to fragmentation events small particles are constantly replenished. But the result on the right side of 
proportionality \eqref{eq:ptesprop} does not, in principal,
depend on whether growth is limited by drift or fragmentation---as long as the conditions for a 
radius-independent pebble flux are fulfilled. Furthermore, as can be seen in Fig.~\ref{fig:pebflux}, the pebble flux 
may only vary with $r$ within roughly one order of magnitude, which is still good enough to estimate the slope of $\surfp$.

\section{Discussion}\label{sec:discussion}
\subsection{Limitations of the current model}
In this paper, the gas density is kept constant, \ie, we neither allow the disk to evolve viscously \citep{Luest1952,Pringle1981} nor do we include sink terms such as photoevaporation 
\citep[{\eg}][]{owen2012}. We also don't include a disk build up phase as tested 
by \cite{birnstiel2010} or \cite{drazkowska2018}. In order to
isolate the influence of the new parameters of the model for analysis, we assume a constant temperature profile according to Eq.~\eqref{eq:Tprofile} and 
ignore viscous heating \citep{Pringle1981}. %as well as time evolution of the stars luminosity. 
We keep $\vdrift$ as for a smooth gas profile, but regions with shallow pressure profiles or even pressure bumps would slow down 
the drift speed. This effect was not considered here since we have a lack of information
about how exactly $\vdrift$ would be 
affected by zonal flows, vortices, or other structures leading to a significant slow-down of particles and subsequent 
gravoturbulent planetesimal formation. 
Also, planetesimals stay in the annulus where they are born. Their mass and radius are not allowed 
to change, \ie, we ignore collisions between them and also growth due to pebble accretion \citep{OrmelKlahr2010}, as well as 
radial motion. Furthermore, we can only give a first impression of how planetesimal formation 
is described and behaves in this model, since we have not explored the full parameter space ($\Mdisk$, $\rchar$, $\dtog(t_0)$,...). 

The turbulence parameter $\aturb$ mostly controls the relative velocities between grains 
\citep{brauer2008coagulation}. Though
this parameter from Eq.~\eqref{eq:turbvisc} can
be given by 
disk winds which cause the transport of angular momentum \citep[see \eg][]{PapaloizouLin1995,bethune2017}, it can also be entangled in, \eg, vortex formation. Hence, a trap formation time, together with a $\aturb(r,t)$ model, 
would make the results more realistic---especially in the outer disk region where it may take a long time to form vortices. Due to conservation of momentum, there is not only the 
influence of gas on dust but also a back-reaction of particles on gas \citep{tanaka2005,Nakagawa1986}. 
This back-reaction becomes
important only for $\dtog\gtrsim0.1$, which is not reached in the simulations 
presented in this paper. But without planetesimal formation, this value can be reached in the vicinity of the ice line where particles are forced to fragment and, thus, to slow down. 
The sink term of pebbles due to planetesimal formation confines this effect. 
Furthermore, we do not consider the bouncing or charging barrier and we do not trace volatiles.
% \chris{resolution? traps resolved? local structure? see comments in the beginning of document}

As pointed
out by \cite{drazkowska2014}, Smoluchowski solvers need a high resolution in mass to mimic the upper end of the size distribution well enough. Especially in the high $\aturb=0.01$ case this upper end can be important for planetesimals formation.

\subsection{Comparison to other models}
In this section we will compare with the most prominent models 
for planetesimal formation. 
\subsubsection{Continuous fluffy growth}\label{sec:fluffy}
% \begin{itemize}
% \item fragmentation barrier \cite{BlumMuench1993}, solved by \cite{wada2007} and \cite{wada2009} (by using ice, small monomers)
% \item radial drift barrier Adachi 1976, solved \cite{okuzumi2012}
% \item compactification problem \cite{suyama2012}, solved \cite{kataoka2013}
% \item works up to 10 AU in a MMSN \citep{okuzumi2012} where the growth timescale is smaller than the drift timescale but only outside the snow line where ice causes high threshold breakup speeds
% \end{itemize}
\cite{okuzumi2012} and \cite{kataoka2013} describe continuous growth to planetesimals. 
They assume that the threshold velocity for fragmentation is never reached, thus avoiding disruptive events. Without replenishment 
of small particles these shouldn't be seen in observations. This "fluffy path" of planetesimal formation is highly sensitive to the 
disk parameters ({\eg} $\aturb$) and works roughly up to $10\,\AU$ for the conditions presented in these papers. Another problem would be that the 
size distribution of initial planetesimals cannot explain the kink feature we see in the size distribution of the Asteroid belt \citep{bottke2005}, the Kuiper belt \citep{FuentesHolman2008,FraserKavelaars2008}, Jupiter trojans \citep{Jewitt2000}, and of Neptune trojans \citep[see their Fig.~4 for an overview]{SheppardTrujillo2010} today. Even though runaway growth may start at the size of the kink feature
\citep{kobayashi2016} which depends on the turbulence level $\aturb$ and the mass which is initially available (see their Fig.~6). 
But this feature is covered by the theory of planetesimal formation via gravitational instability \citep{morbidelli2009,KlahrSchreiber2015,andydiss}, almost independent of disk radius and very robust against changes in $\aturb$. 
However, one cannot dismiss the idea of fluffy growth entirely, especially in the early stages of particle growth. Even some planetesimal could be formed this way without a conflict with today's data. 

However, as shown by \cite{krijt2015}, erosion, where particles lose mass due to collisions with smaller projectiles, may prevent growth through the radial drift barrier. They also show that only for high erosion velocity values of $60\,\mathrm{m/s}$ porous particles can overcome this barrier within $10\,\AU$ for a minimum mass solar nebula \citep{weidenschilling1977mmsn,Hayashi1981}.
\subsubsection{Lucky particle growth}\label{sec:lucky}
By including a Maxwell-Boltzmann like velocity distribution, lucky particle effects can occur as described by \cite{windmark2012} and tested by 
\cite{drazkowska2014proceeding} and \cite{estrada2016}. 
% \chris{booth2017 has also shown that the bouncing barrier can be overcome in the case without radial drift}. 
{\Ie}, growth barriers are smoothed out or can be overcome by low velocity collisions leading to growth where the mean 
collisional speed would lead to 
bouncing or fragmentation. Although these low velocity collisions have a low probability, by experiencing multiple lucky collisions 
in a row, the growth barriers can be overcome. In the case of the bouncing barrier, this effect is even stronger because 
higher collision velocities lead only to bouncing, {\ie} no disruption but possible further compactification, but low velocity 
collisions let the grains grow. Therefore, the bouncing barrier may not be a strict solid barrier as long as the growth timescale (considering bouncing)
is sufficiently small. 
By overcoming the bouncing and fragmentation barrier, these particles present the seeds for a 
sweep-up scenario which can cause a growth process toward planetesimal size. 

But this path is quite inefficient in forming planetesimals and it did not take drift into account. 
% The fragmentation barrier is roughly symmetric around $\stokes=1$. To grow from Stokes numbers of $0.1$ to $10$ one needs to grow 
% in the Epstein regime ($\stokes\propto\dustrad\propto m^{1/3}$) by a factor of $100$ which means a factor of $10^6$ in mass. 
% If 
% the lucky particles grow via sweep-up of the particles at the fragmentation barrier, say $\stokes=0.1$ particles, one would need 
% $10^6-1$ lucky collisions without a single disruptive collision. If many particles would be lucky enough to grow monodispersely, this 
% would still take $20$ ($10^6m=m\cdot2^x$) of such non-destructive collisions leading to a doubling in mass. 
The key here is to have most collision partners leading to mass transfer or sticking. But a lower fragmentation threshold speed 
within the ice line would basically make it impossible to grow beyond the barriers (or very unlikely). And in the outer disk, drift 
is removing the lucky particles faster than they can grow beyond the bouncing barrier \citep{drazkowska2014proceeding,estrada2016}.

\subsubsection{Planetesimal formation directly regulated via streaming instability}\label{sec:SImodel}
\cite{drazkowska2014} and \cite{drazkowska2016} proposed a parameterized formation model of planetesimals by asking when and where
the pebble density ($\stokes>0.01$) in the disk midplane reaches a critical value of $\rhod/\rhog=1$, assuming a balance of sedimentation and vertical diffusion of pebbles. %at an efficiency of $\zeta=10^{-4}$. 
At this critical dust-to-gas ratio it can be argued that the streaming instability will be triggered \citep{YoudinGoodman2005}. 
% \chris{\cite{carrera2017} found that a midplane dust-to-gas ratio is the stronger condition compared to the necessary enhancement in $Z$, see their figure 2.}
But as was shown by \cite{johansen2009}, \cite{Carrera2015}, and \cite{yang2017} 
triggering the streaming instability at 
$\rhod(\stokes>10^{-2})/\rhog=1$ is not sufficient to create strong enough over-densities
to lead to gravitational collapse.
The additional criterion to have the total local column density dust-to-gas ratio ($\dustsurf/\surfgas$) increased by a factor of 2--4 seems to be implicitly tested by 
having sufficient sedimentation even at $\aturb=10^{-4}$ as vertical diffusion. But, as argued by \cite{drazkowska2016}, and also shown by \citet[see their Fig. 2]{carrera2017}, reaching the critical
midplane dust-to-gas ratio is typically the stronger condition. Similar to our ansatz, \cite{drazkowska2016} convert pebbles 
into planetesimals with a certain efficiency per orbit, where the column density formation rate 
$\dotsurfp\propto\dustsurf(\stokes>10^{-2})$ in their model. {\Ie}, they use a local approach whereas we focus on the pebble delivery via the local pebble flux.
 
In the work of the follow-up paper by \cite{drazkowska2017}, they follow a trap that will actually show up in 1D radial disk models, e.g. pebble pile beyond the (water) ice line due to traffic jam and recondensation of water vapor. This means planetesimal formation will mainly happen outside of and close to the ice line in their model. As a result, planetesimals will be confined to a relatively narrow region between 2 and 5$\,\AU$. Besides that they likely will be too water rich to explain the composition of terrestrial planets. 
Not only are planetesimals only formed where the water ice line is moving (relatively small annulus compared to the disk size), but the time for planetesimal formation is also very limited ($\sim2-3\cdot10^5\,\yr$). 
% A detailed comparison of the therein predicted column density of planetesimals will be given in the discussion section.
% \begin{itemize}
% \item $\rho_\mathrm{d}(z=0,\stokes)/\rho_\mathrm{g}(z=0)=\dustsurf(\stokes)/\surfgas\cdot\sqrt{1+\stokes/\aturb}$
% \item $\sum_{\stokes>10^{-2}}\rho_\mathrm{d}(z=0,\stokes)/\rho_\mathrm{g}(z=0)>1$?
% \item $\dotsurfp=\zeta \dustsurf(\stokes>10^{-2})/\torb$
% \end{itemize}
% Where $\zeta$ is her planetesimal formation efficiency per orbit, or in other words, the ratio of pebble mass turned into planetesimals within one orbit.
\begin{table*}[t]
\caption{Overview of different approaches leading to streaming instability (SI) or planetesimals (ptes). With 
$\aturb=0$ we mean that there is no turbulence active, or only turbulence
stemming from streaming instability itself.
% With $\aturb$ we mean that this turbulence parameter was set due to initialization and not the value 
% that would develop during the simulation caused by {\eg} streaming instability. 
The second quantity is the
sub-Keplerianness, defined as 
$\eta:=1/2\cdot(\hscale/r)^2\cdot\dif{\ln{\rhog}}/\dif{\ln{r}}$. In the third column with "feedback" we mean particle-gas coupling such that momentum is also transferred to the gas if particles loose momentum due to the interaction with the gas. 
The velocity dispersion of particles is $\vdisp$, $\Okepl$ is the Kepler frequency, and $\Grav$ the gravitational constant. For larger particles $\vdisp$ decreases, allowing gravitational instability for large particles \citep{weidenschilling1995}. 
The question mark in the third row indicates that one cannot be entirely sure that streaming instability is active.
} 
\label{tab:SIoverview}
\centering
\begin{tabular}{cccc|cc|c}
 \toprule\toprule
 \multicolumn{4}{c}{input} & \multicolumn{2}{c}{result} & 
 \\
 $\aturb$  	& $\eta$    & feedback  & $Z=\dustsurf/\surfgas$        & SI    & ptes  &  references
 \\
 \midrule
 $=0$       & $=0$      & no        & $>\Okepl\vdisp/(\pi\Grav\surfgas)$        & no    & yes   &  \cite{Safronov1969,GoldreichWard1973,YoudinShu2002} 
 \\
 $>0$       & $>0$      & no        & 0.01       & no    & yes   & \cite{Johansen2006}
 \\
 $>0$       & $>0$      & yes       & 0.01       & ?     & yes   & \cite{johansen2007nat}
 \\
 $=0$       & $>0$      & yes       & 0.01       & yes   & no    & \cite{johansen2007nat} 
 \\
 $=0$       & $>0$      & yes       & $\gtrsim0.03$ & yes   & yes   & \cite{johansen2009,simon2016}; \citet[$Z=0.1$, $\eta$ varies]{abod2018} \\
 \bottomrule
 %\bottomrule
\end{tabular}
\end{table*}

\begin{figure}[tb]
  \centering
  \includegraphics[width=.47\textwidth]{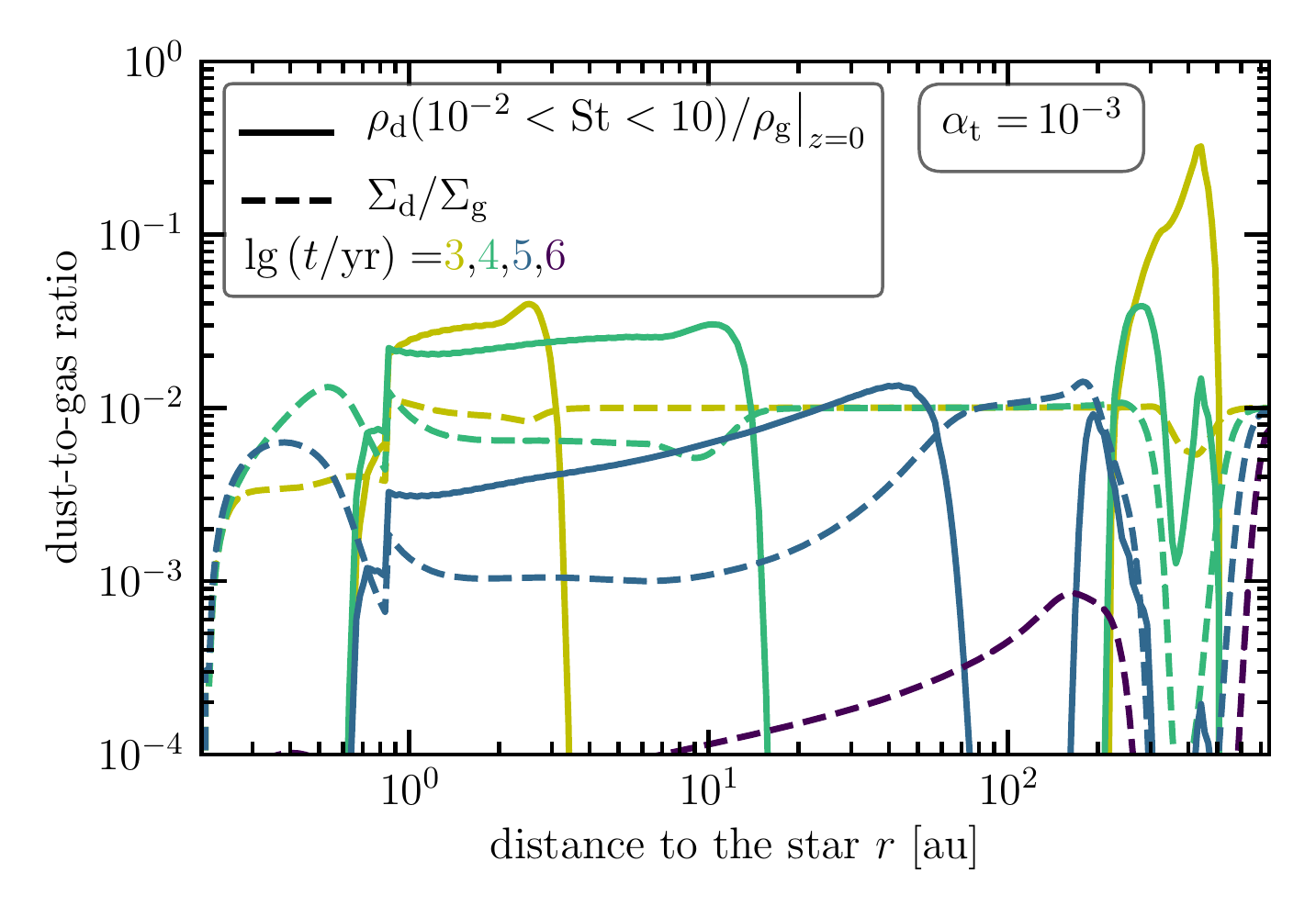}
  \caption
  {Snapshots from $10^3\,\yr$ (yellow) to $10^6\,\yr$ (purple) of the dust-to-gas ratio in the disk midplane of $10^{-2}<\stokes<10$ particles (solid) and the column density ratio $\dtog=\dustsurf/\surfgas$ of all particles (dashed). The turbulence level is $\aturb=10^{-3}$. Though 
  the dust scale height from \cite{Dubrulle1995}, see Eq.~\eqref{eq:hscdust}, is only valid for $\stokes<1$ particles, to follow
  and compare to \cite{drazkowska2016} we adopted this expression also for particles with larger Stokes numbers (in the outer 
  disk). Particles close to the outer edge of the disk can reach large Stokes numbers $>1$ due to the low gas density. 
  At $10^6\,\yr$ the midplane dust-to-gas ratio is smaller than $\dtog$ since most of the mass is not in $10^{-2}<\stokes<10$ 
  particles anymore.
   }
  \label{fig:d2g}
\end{figure}

Since in our model planetesimals are also formed in the outer disk, less material will reach the inner part. Hence, 
we may not reach their condition $\rhod(\stokes>10^{-2})/\rhog\geq1$, which we illustrate in Fig.~\ref{fig:d2g}. Even though 
the back-reaction of particles on gas and the effect of collective drift may not be important here, tracing water vapor could make a difference. Nevertheless, 
we don't reach the condition used by \cite{drazkowska2016} since the midplane dust-to-gas ratio is more than one order of magnitude below the critical value of unity over 
almost the entire disk. 
% Even though analytical calculations indicate that streaming instability may 
% not work outside of pressure bumps \citep{auffinger2017}, 
In principle, a combined formation model of both pebble flux regulated formation 
in pressure bumps and the local conversion could coexist. Maybe photoevaporation can reduce the gas content at late times 
($\gtrsim10^6\,\yr$) so much that the condition can be reached in some parts of the disk.

We would like to emphasize that streaming instability per se is \emph{not} mandatory for gravitational assisted planetesimal 
formation, and in fact it prevents planetesimal formation for
low enough overall dust-to-gas ratios in the disk \citep{johansen2007nat}. 
As in star formation \citep{klessen2005}, turbulence plays a dual role. On global scales it promotes planetesimal
formation by 
leading to clumps and enhancing the local dust-to-gas ratio, while at the same time it can prevent collapse on smaller scales by diffusing material 
away. \Ie, turbulence stabilizes dust against collapse on small scales \citep{andydiss}. Also the resulting 
planetesimal size is a question of particle diffusion, which does not necessarily have to stem from 
streaming instability. 
We summarize the work that has been done on fundamental research concerning planetesimal 
formation with and without turbulence in Table~\ref{tab:SIoverview}. If disk models assume some non-zero value for $\aturb$ as in \cite{drazkowska2016}, it means that large scale turbulence is active. Whether there is a critical $Z$, $\rhod/\rhog$
or streaming instability 
criterion for these turbulent cases (magneto-rotational instability (MRI) etc.) in order to form planetesimals
has not been studied so far. 
The simulations by \cite{johansen2007nat} and \cite{johansen2011} 
were in fact for MRI turbulent cases, and found planetesimal formation in zonal flows with streaming instability playing a minor role in 
enhancing perturbations on smaller scales. Planetesimals
formed with and without particle feedback, {\ie}, streaming instability. These 
simulations unfortunately had to use quite large particles ($\stokes=0.25$), which is due to the problem of covering the large 
scales of MRI turbulence on scales of $\sim\hscale$, whereas the streaming instability unstable wavelength for smaller Stokes number particles of $\stokes = 0.1-0.01$ is 
more like $0.01\hscale$. This makes 3D simulations of MRI and 
streaming instability leading to planetesimal formation for 
small Stokes numbers too challenging at the moment.

Higher $Z$ damps streaming instability \citep{andydiss}, which means lower diffusion. And when there is less 
diffusion, planetesimals form (\ie, for weaker streaming instability), as seen in \citet[see their table 2 and their section 7.2.4.]{BaiStone2010}. In 
our Table~\ref{tab:SIoverview} we collect evidence that streaming instability is not needed for planetesimal formation, but it is a controlling effect---it can even prevent the formation. Hence, we suggest
following \cite{Johansen2006} and calling the process "gravoturbulent planetesimal formation". 
Remember, in a disk without a local radial pressure gradient, there is no streaming instability and the \cite{GoldreichWard1973} mechanism will work perfectly as recently demonstrated by \cite{abod2018}.

% \cite{johansen2009} have shown that ratios of $\dustsurf/\surfgas>(2-3)\cdot10^{-2}$ are enough to trigger streaming instability. But this also depends on 
% $\aturb$. 
% Furthermore, analytical calculations by \cite{auffinger2017} show that the classical streaming instability 
% in the $r$-$z$ (radial-vertical) plane prefer pressure bumps to be active. But there exists also streaming 
% instability in the $r$-$\varphi$ (radial-azimuthal) plane \citep{SchreiberKlahr2018}.
\cite{auffinger2017} find streaming instability to be active in local pressure maxima, despite the pressure gradient
vanishing locally. Whether this triggering of streaming instability has an impact on the formation of planetesimals has not been studied in detail. Yet,
\cite{johansen2007nat} and \cite{johansen2011} find indeed the formation of planetesimals in these bumps, leaving the role of
streaming instability undetermined so far. Furthermore, there can also be streaming instability in the $r$-$\varphi$ (radial-azimuthal) plane \citep{SchreiberKlahr2018}.

\section{Conclusion and Outlook}\label{sec:concl}
We presented a new model for the planetesimal formation rate that is directly linked to the local pebble flux---and not local 
dust-to-gas ratio or local density---relying on parameters such as lifetime of pressure-bump structures, 
contributing particle Stokes numbers ($\stmin$ \& $\stmax$), the radial particle trap density ($1/\trapdist$), and an 
efficiency parameter $\peff$ describing the conversion from pebbles to planetesimals. We have implemented 
the presented model into a sophisticated dust and gas evolution code, where we switched off gas evolution. This model was investigated for two different 
values of the turbulence parameter, $\aturb=10^{-3}$ (moderate turbulence) and $10^{-2}$ (strong turbulence), as well as 
two different efficiency values, $\peff=0.1$ and $0.8$. The results can be summarized as follows:
\begin{itemize}
\item This model is the first one creating planetesimals \emph{everywhere} in the disk from a few $\AU$ to a few hundred $\AU$.
\item The resulting planetesimal distribution and the total mass of initial planetesimals highly depends on the level of turbulence ($\aturb$ value), {\ie} whether particle growth 
is limited by fragmentation or drift. 
\item If the disk is fragmentation limited, higher $\peff/\trapdist$ leads to
higher planetesimal formation rates, but these also decline faster. Nevertheless, higher efficiencies yield more planetesimals. 
\item %\chris{if the planetesimals are formed in a drift limited disk} 
If the disk is mostly limited by drift, material drifts faster and particles which were not converted into 
planetesimals in the outer disk can build planetesimals in the inner part. That is, if the level of turbulence is not too 
high (here moderate turbulence for $\aturb=10^{-3}$), the slope of the 
final planetesimal profile can be steeper than that of initial dust and gas. 
The overall planetesimal profile $\surfp$ looks also more like a power-law in the inner disk part, with an exponential roll 
off in the outer part for lower turbulence.
\item A few planetesimals should have formed in the terrestrial region. Thus, strong turbulence ($\aturb=10^{-2}$) seems to be unlikely for the Solar Nebula in order to form the Solar System, 
especially Earth---
at least as long as the minimum Stokes number contributing to planetesimal formation is as high as $\stmin=10^{-2}$ 
and the efficiency parameter $\peff$ is not very high ($\gtrsim0.8$ for $\trapdist=5\hscale$).
\item The late stage planetesimal-to-dust ratios indicate a larger feeding zone of planetesimals, 
{\ie} material forming planetesimals can originate from far
away if $\peff/\trapdist$ is small. Here, it is 
difficult to tell how small exactly, since the feeding zone will stretch with smaller values of that ratio. 
In this paper, $0.1/(5\hscale)$ is small compared to $0.8/(5\hscale)$. 
But one has to investigate in the future how exactly this feeding zone scales with $\peff/\trapdist$.
\end{itemize}

Future work will analyze the influence of disk evolution and other disk parameters such as the initial disk mass and disk size. Implementing of a proper temperature model with accretion heating, photoevaporation, planetesimal collisions, and pebble accretion will be the focus of future work. 
For an upcoming study, one could
compare the outcome of our simulations for the initial planetesimal population with constraints in special regions of the solar nebula to exclude 
certain pairings of disk parameters, or even extreme cases for specific parameters such as $\aturb\geq 10^{-2}$.
The efficiency parameter $\peff$ of the combination of trapping and planetesimal formation can depend on both the particle Stokes number and the disk radius. Fluid dynamics codes have to show how this efficiency depends on particle and disk properties. 
In this paper, we assumed that particle traps are present from the beginning of the simulation. In reality these traps will take 
some time to form. Thus, one could introduce another parameter which describes the trap formation time. 
During this time, particles can drift inward, reducing the mass reservoir and the pebble flux available for 
transformation into planetesimals once particle traps are considered to be active.

\acknowledgments
C.L. is thankful to Andreas Schreiber, Joanna Dr{\k{a}}{\.z}kowska, Cornelis Dullemond, Akimasa Kataoka, Joachim A. Maruhn, Andrew Youdin, Konstantin Gerbig, Katherine Kretke, Hans Baehr, and Vincent Carpenter for helpful discussions. 
We also thank the anonymous referee for a useful report. 
This work was funded by the Deutsche Forschungsgemeinschaft (DFG,  German  Research  Foundation) as part of the Schwerpunktprogramm (SPP, Priority Program) SPP 1833 
``Building a Habitable Earth'' and in part  at KITP Santa Barbara by the National Science Foundation under Grant No. NSF PHY17-48958. Part of this work was performed at the Aspen Center for Physics, which is supported by National Science Foundation grant PHY-1607761. This research was supported by the Munich Institute for Astro- and Particle Physics (MIAPP) of the DFG cluster of excellence 
``Origin and Structure of the Universe''. T.B. acknowledges funding from the 
European Research Council (ERC) under the European Union's Horizon 2020 research and innovation program under grand agreement No. 714769 and funding from the DFG Ref no. FOR 2634/1.
\begin{appendices}
  \appendix
  \section{The coagulation approach}\label{sec:coag_app}
  The first mathematical expression (in discrete form) for the coagulation process was obtained by \cite{Smoluchowski1916}. It was first written 
in integral form by \cite{Schumann1940}, which is appropriate for a continuous mass spectrum. 
Consider a dust grain distribution $\nm(\msol,r,z)$ 
%($[\nm]=\mathrm{cm^{-3}\,g^{-1}}$) 
as the number of particles per spatial volume and per particle mass interval $[\msol,\msol+\dif\msol]$. In protoplanetary disks, $\nm$ is a function of the 
solid particle mass $\msol$ \emph{as well as} the distance $r$ and height $z$ (cylindrical coordinates). Since we assume a cylindrically symmetric 
system, the quantities do not depend on the azimuthal angle $\varphi$. 
Thus, the total dust mass density 
%($[\rhod]=\mathrm{g/cm^{3}}$) 
is
\begin{equation}
\label{eq:rhod-smolu}
 \rhod(r,z)=\int_0^\infty \dif\msol\ \nm(\msol,r,z)\cdot\msol.
\end{equation}
A further generalized coagulation equation including fragmentation would be \citep[see \eg,][]{barrow1981,birnstiel2010}
\begin{subequations}
\begin{align}
  \begin{aligned}
  \label{eq:Smoluchowskiwithfragments}
  \pd{}{t}\nm(\msol,t,r,z)=&\int_0^{\infty}\dif\msol_1\int_0^{\infty}\dif\msol_2\ K(\msol,\msol_1,\msol_2,r,z)\\
			   &\times\nm(\msol_1,r,z)\nm(\msol_2,r,z),
  \end{aligned}
\end{align}
where the reaction kernel is
given by
\begin{equation}
  \begin{aligned}
    \label{eq:kernelmitfrag}
    K(\msol,\msol_1,\msol_2):=\; &\frac{1}{2}\coagkern(\msol_1,\msol_2)\cdot\DD{\msol_1+\msol_2-\msol}\\
    &-\coagkern(\msol_1,\msol_2)\cdot\DD{\msol_2-\msol}\\
    &+\frac{1}{2}\fragkern(\msol_1,\msol_2)\cdot \fragdistr(\msol,\msol_1,\msol_2)\\
    &-\fragkern(\msol_1,\msol_2)\cdot\DD{\msol-\msol_2}.
  \end{aligned}
\end{equation}
\end{subequations} 
In this expression $$\coagkern(\msol_1,\msol_2):=\cprob(\msol_1,\msol_2,\vrelcoll)\cdot\vrelcoll(\msol_1,\msol_2)\cdot\crossseccoll(\msol_1,\msol_2)$$ 
is the coagulation kernel, 
$$\fragkern(\msol_1,\msol_2):=\fprob(\msol_1,\msol_2,\vrelcoll)\cdot\vrelcoll(\msol_1,\msol_2)\cdot\crossseccoll(\msol_1,\msol_2)$$  the fragmentation kernel,
and $$\crossseccoll(\msol_1,\msol_2)=\pi (\dustrad_1+\dustrad_2)^2$$ the geometrical cross section , where $\dustrad_{i}$ is the particle radius of the colliding particle with mass $\msol_i$.
$\cprob$ and $\fprob$ denote the probability that a collision between two particles of mass $\msol_1$ and $\msol_2$ with a relative velocity $\vrelcoll$ at collision leads to coagulation or fragmentation ($\cprob+\fprob=1$), respectively. 
Colliding particles with $\msol_1$ and $\msol_2$ give a gain term for particles with mass $\msol$, determined by 
the distribution of fragments $\fragdistr(\msol,\msol_1,\msol_2)$, where we assume that the 
distribution is a power-law according to
\begin{equation}
	\label{eq:frdist}
	\nm\dif{\msol}\propto\msol^{-\frd}\dif{\msol}.
\end{equation}
We follow \cite{brauer2008coagulation} and use $\frd=1.83$, but this power-law index may depend on the collision speed 
\citep[his section 4.2]{HusmannDiss}. 
% \chris{The overall result does not change significantly for different values of $\frd$ as 
% long as it is not changed ba a factor greater than $2$ as indicated by \cite{birnstiel2011} (see their Fig.~2). what about Brauer diss Fig.~5.9?} 
The Dirac delta-distribution is denoted by $\DD{\cdot}$.

The first term in Eq.~\eqref{eq:kernelmitfrag} corresponds to an increasing number of particles with mass $\msol$ 
due to grain growth, \ie, particles with masses $\msol_1$ and $\msol-\msol_1$ coagulate. The second term represents the loss of
particles with mass $\msol$ ($\msol_1$ coagulates with $\msol$). The third term stands for the fragmentation due to a 
collision between particles with mass $\msol_1$ and $\msol_2$, and includes the fact that subsequently a distribution of some of their 
mass will occur via fragmentation to smaller sizes. The fourth term describes the fragmentation of particles with mass $\msol_1$ and $\msol$ 
and thus represents the \emph{loss} of particles with mass $\msol$. 
The factors $1/2$ eliminate double counting of the collisions increasing the number of particles of mass $\msol$. We use 27 bins per size decade. The radial grid has 300 points spanning from $0.2$ to $800\,\AU$. The grid points are
logarithmically equally spaced, both in mass and disk radius. In order to optimize computational time, we only numerically consider mass cells up to twice the mass of the current maximum mass in the system at each disk radius, similar to section 2.2 of \cite{lee2000}.

In general, all these quantities can also depend on other material properties such as composition, porosity and charge. To keep it simple we consider just the mass 
of the colliding particles. As Eq.~\eqref{eq:Smoluchowskiwithfragments} is proportional to the product of two densities and particle densities become larger toward the mid-plane due to 
vertical settling, we follow \cite{birnstiel2010} and approximate the kernels by the mid-plane values. However, the 
$z$-dependence is eliminated by vertical integration, assuming a Gaussian $z$-profile for the particle density. In the code, the \emph{discrete} Smoluchowski equation is solved numerically. 
See Appendix A of \cite{birnstiel2010} for details. 
  \begin{figure*}[ht]
      \centering
      \includegraphics[width=0.8\textwidth]{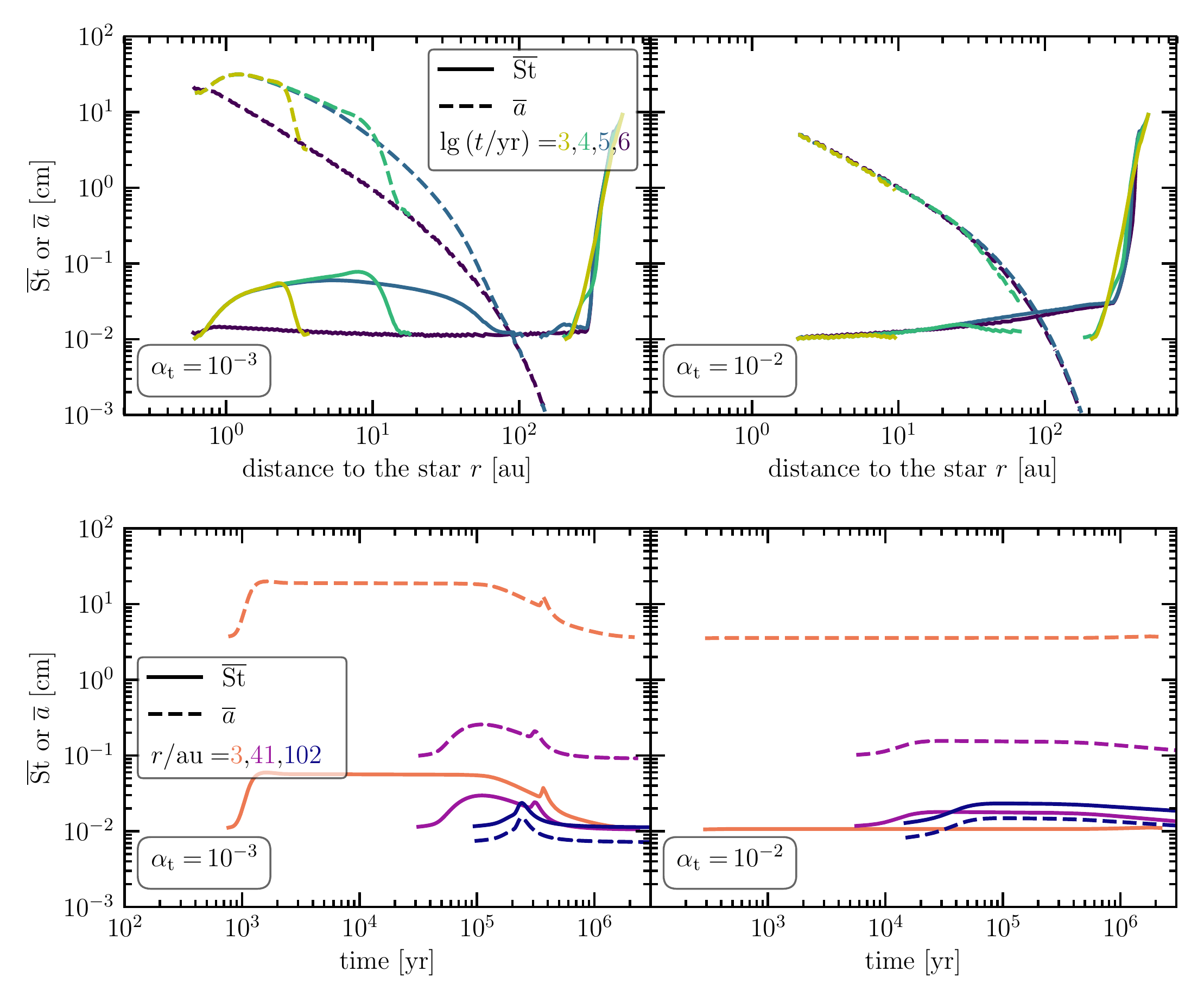}
      \caption
      {Evolution of the flux averaged Stokes numbers in the mid-plane (solid) and particle radii (dashed) for a simulation with $\aturb=10^{-3}$ (left panels) and $\aturb=10^{-2}$ 
      (right panels). Both are shown as a time series as a function of radius (top panels) and for different radii as a function of time (bottom panels). 
      In all plots the left axis belongs to $\stav$ and the right one to $\dustradav$ according to the definition in Eq.~\eqref{eq:fluxav}. 
%       \chris{2 plots for $\stokes$ and $\dustrad$. For the former only show values up to $\stokes=10^{0}$ or even 
%       $10^{-1}$ to zoom in. For the $\stokes$ plot one could add the average value where the flux is averaged over all particles, not
%       only $\stmin<\stokes<\stmax$. This can then also be used to compare to the twopoppy model, {\ie} with the growth barriers 
%       including the fudge factors.}
      }
      \label{fig:St_av_a_av_setup01_02}
    \end{figure*}
  \section{Radial Particle velocities dominated by gas flow}\label{parvel}
   It is worthwhile to estimate the Stokes number below, which the first velocity 
    term in Eq.~\eqref{eq:vdusttotal} dominates. Therefor we use 
    \begin{align}
     \label{eq:hscaleprop}
     \hscale\propto r^{(3-\expT)/2},
    \end{align}
    assuming a falling temperature power-law with exponent $\expT$. Further, we assume 
    a gas column density profile following Eq.~\eqref{eq:gasini} with $\vispow=1$, 
    $\rchar=35\,\AU$. This leads to the logarithmic pressure gradient
    \begin{align}
     \label{eq:dfracpd}
     \dfracpd{\ln{\press}}{\ln{r}}=-\vispow-\dfrac{3+\expT}{2}-(2-\vispow)
        \left(\dfrac{r}{\rchar}\right)^{2-\vispow}
    \end{align}
    which is obtained by using $\press=\sound^2\rhog$ for mid-plane values, see Eq.~\eqref{eq:rhozerogas}. 
    There are regions in the $r$-$\stokes$ space where the total particle velocity is 
    dominated by the gas velocity, {\ie}, by the left term in Eq.~\eqref{eq:vdusttotal}. 
    We will now determine the condition 
    on the Stokes number for this term to dominate the radial velocity. 
    % in order that the radial velocity is dominated by this term. 
    In the most cases the dust-to-gas ratio is small on large scales, thus, we assume $\rhod/\rhog\ll1$. This way we obtain
    \begin{align}
      \frac{3}{\surfgas\sqrt{r}}\abs{\pd{}{r}(\surfgas\turbvisc\sqrt{r})}\geq\stokes\frac{\hscale}{r}\abs{\pd{\ln{\press}}{\ln{r}}}\sound
    \end{align}
    which can be written in the form 
    \begin{align}
     \frac{3}{r}\turbvisc\abs{\pd{}{\ln{r}}\ln{(\surfgas\turbvisc\sqrt{r})}}\geq\stokes\frac{\hscale}{r}\abs{\pd{\ln{\press}}{\ln{r}}}\sound
    \end{align}
    By inserting the turbulent viscosity from Eq.~\eqref{eq:turbvisc} as well as Eqns.~\eqref{eq:hscaleprop} and \eqref{eq:dfracpd}
    we get, for a constant $\aturb$ (after rearranging)
    \begin{align}
     \nonumber
     &3\aturb\abs{-\vispow-(2-\vispow)\left(\dfrac{r}{\rchar}\right)^{2-\vispow}+\frac{3-\expT}{2}-\frac{\expT}{2}+\frac{1}{2}}
      \\
      \geq&\,
      \stokes\left[\vispow+\dfrac{3+\expT}{2}+(2-\vispow)\left(\dfrac{r}{\rchar}\right)^{2-\vispow}\right]
    \end{align}
    which finally yields Eq.~\eqref{eq:stgasdom}. 
    If $\rhod/\rhog\ll1$ is not fulfilled, this value has to be multiplied
    by $[1+(\rhod/\rhog)^2]$ for mid-plane values. 
    Since $\surfgas$ decreases exponentially for $r/\rchar>1$, whereas $\turbvisc\sqrt{r}$ increases with $r$, there exists a radius from which gas is flowing outward. This outflow of gas can drag dust with
    sufficiently small Stokes numbers along. The transition radius where the gas velocity $\vgask{r}$ switches from an inflow ($\vgask{r}<0$) to an outflow ($\vgask{r}>0$) can be computed 
    the same way:
    \begin{align}
      \nonumber\vgask{r}\overset{!}{\geq}0
    \end{align}
    leads with Eq.~\eqref{eq:ugasr} to
    \begin{align}
      \nonumber
      &\pd{}{\ln{r}}\ln{(\surfgas\aturb\hscale\sound\sqrt{r})}
      \\
      =&\,-\vispow-(2-\vispow)\left(\dfrac{r}{\rchar}\right)^{2-\vispow}+2-\expT
      \leq0%\\ \displaybreak[1],
    \end{align}
    giving
    \begin{align}  
      r\geq\left(\dfrac{2-\expT-\vispow}{2-\vispow}\right)^{1/(2-\vispow)}\rchar.
      \label{eq:negativegasvel}
    \end{align}
    Since in our simulations the pressure gradient is always negative, drift is also always pointing inward. In this case, 
    the region of outflowing solid matter in the $r$-$\stokes$ space is given by Eq.~\eqref{eq:stgasdom}, giving the Stokes number of the velocity sign flip, 
    and Eq.~\eqref{eq:negativegasvel}, giving the disk radius
    beyond which gas, and thus also these particles, are flowing outward.
%   \section{High efficiencies}
%     \begin{figure*}[th]
%       \centering
%       \includegraphics[width=.75\textwidth]{s01_e08_r35_M5_flux_St_4_5_6_nodiskev.pdf}
%       \caption
%       {
%       Some super smart caption! $\aturb=10^{-2}$, $\peff=0.8$. At late times (here $10^6\,\mathrm{yr}$) planetesimal formation has reduces the dust and pebble population so much that collisions became a rare event. Thus, they are able to drift inward even though they hit the fragmentation barrier.
%       }
%       \label{fig:s01_e08_r35_M5_flux_St_4_5_6_nodiskev}
%     \end{figure*}
%     
%     \begin{figure*}[th]
%       \centering
%       \includegraphics[width=.75\textwidth]{s02_e08_r35_M5_flux_St_4_5_6_nodiskev.pdf}
%       \caption
%       {
%       Some super smart caption! $\aturb=10^{-3}$, $\peff=0.8$
%       }
%       \label{fig:s02_e08_r35_M5_flux_St_4_5_6_nodiskev}
%     \end{figure*}
    
    \section{Averaged Size of planetesimal forming material}
    What is the average size of the material which is building planetesimals? To answer this question we use flux averaging such 
    that 
    \begin{align}
    \nonumber
    \overline{X}:=&\,\frac{\sum_{\stmin\leq\stokes\leq\stmax}X\abs{\vdrift(r,\stokes)}\dustsurf(r,\stokes)}{\sum_{\stmin\leq\stokes\leq\stmax}\abs{\vdrift(r,\stokes)}\dustsurf(r,\stokes)}
    \\
    \label{eq:fluxav}
    &\times\theta(\pebflux-\Mcr),
    \end{align}
    where $X$ is either Stokes number, $\stokes$, or particle radius, $\dustrad$, and the sum goes over 
    all particles with Stokes number in the range between $\stmin$ and $\stmax$. The Heaviside function reflects 
    the condition~\eqref{eq:cond_ptes}. We use a flux average, since in our model 
    the contribution to planetesimals scales with it. Fig.~\ref{fig:St_av_a_av_setup01_02} shows that the flux averaged particle 
    size building planetesimals is roughly a constant over time, as long as the dust-to-gas ratio is high enough to keep the drift 
    limit close to the fragmentation barrier. 
    The large peak beyond $300\,\AU$ stems from the exponential drop in gas density. This yields planetesimals which are
    built from $\sim\mathrm{\micron}$-sized particles. The
    $r$-dependence of $\dustradav$ is mostly determined by the gas column 
    density $\surfgas$, as can be seen from the almost $r$-independent $\stav$ values at $10^6\,\yr$, which transforms $\stav$ via 
    Eq.~\eqref{eq:StEp} into $\dustradav\propto\surfgas$.
    
  \section{Necessity of $\stmin$ and $\stmax$}
    If the flux is not dominated by particles in the range $\stmin\leq\stokes\leq\stmax$, these parameters become very important. 
    Though, $\stmax$ is only important for the outer disk part where the gas density becomes very low, leading to high Stokes 
    numbers $>1$ even for the smallest particles in the simulation ($0.1\,\mathrm{\micron}$). Fluid dynamical simulations 
    have to show which values of the Stokes number represent the boundaries at which \emph{both} particle trapping and planetesimal 
    formation can occur in a gravo-turbulent scenario. One prerequisite would be the onset of streaming instability which can 
    further regulate planetesimal formation after enough mass has been accumulated. In Fig.~\ref{fig:pebflux} we plot 
    both the particle flux fulfilling $10^{-2}=\stmin\leq\stokes\leq\stmax=10$ and the flux of all particles. In the 
    extreme case where $\stmin=0$ and $\stmax=\infty$, the difference between the total flux (dashed lines) and 
    the pebble flux shown (solid lines, $\stmin=10^{-2}$, $\stmax=10$) would also yield a difference in planetesimals. 
    Since the
    particles' contribution to planetesimals is weighted by the contribution of their flux to the overall pebble flux, smaller particles also contribute less. Smaller values of $\stmin$, say $10^{-3}$, would lead to 
    earlier planetesimal formation, which also persists longer. The edges of the planetesimal zone would also stretch out. 
    
  \section{Total dust and planetesimal mass in the disk}\label{sec:totdiskmass}
\begin{figure}[t]
  \centering
  \includegraphics[width=0.49
\textwidth]{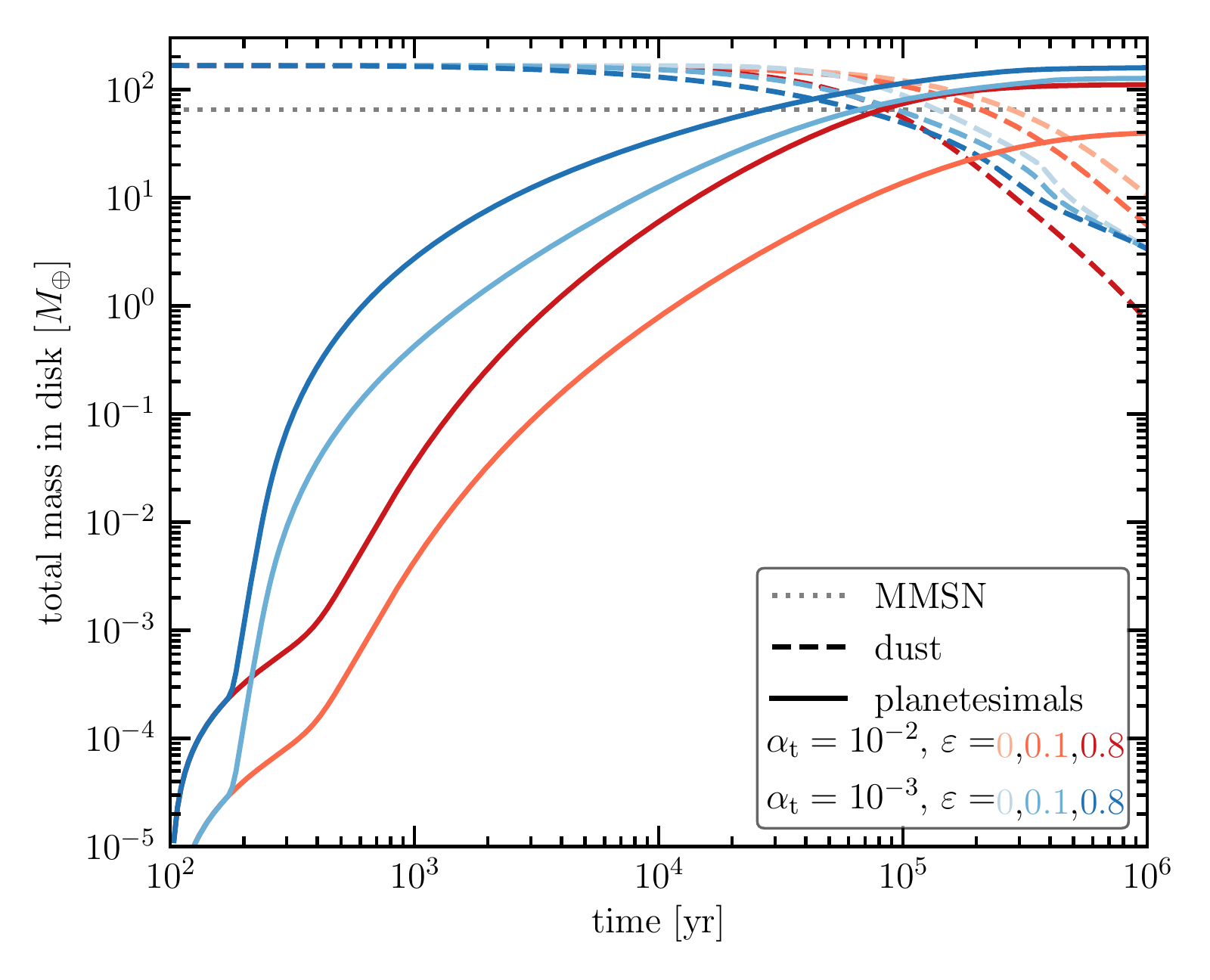}
  \caption
  {
   Total planetesimal mass (solid lines) and total mass of solids not in planetesimals (dashed lines) 
   as a function of time. The gray dotted line shows 
   the total mass in solids of the minimum mass Solar Nebula (MMSN), after \cite{weidenschilling1977mmsn} and 
   \cite{Hayashi1981}. For this plot the same parameters 
   were used as in Fig.~\ref{fig:ptes_nodiskev}. 
%    \chris{Show only mass within up to $60\,\AU$? blue dashed lines overlap at late times because during this time all the particles which are by then in planetesimals would have drifted to the inner edge anyway?}
   }
  \label{fig:totdisk_nodiskev}
\end{figure}

  Fig.~\ref{fig:totdisk_nodiskev} shows the integrated mass of planetesimals and smaller particles. 
  Reaching the total solid mass of the MMSN does not necessarily imply that the result is compatible with planetesimal formation in the Solar Nebula. 
 It is more like a necessary condition and, therefore, knowledge about the planetesimal distribution as a function of radius is mandatory.
 For $\aturb=10^{-2}$ values over the MMSN mass can be reached only for high $\peff$ (here 0.8), which 
 is unlikely since the efficiency in the planetesimal formation process itself is already lower \citep[see $\kappa$ in his Table~7.3]{andydiss}.
 
 High $\aturb$ leads to wider regions which are fragmentation limited. Hence, more mass is in small, slow drifting particles. 
 The total mass strongly depends 
 on $\peff$. Whereas for mostly drift limited disks, the mass reservoir in the outer disk is transported to the 
 inner part much faster. Over this long distance, particles which 
 were not transformed into planetesimals in the outer part may turn into one in the inner part. This can lead to similar 
 disk masses of the planetesimal population but with a different $r$-dependent distribution.

\section{Origin of particles beyond growth barriers}\label{origin_part_grt_growth_barriers}
\begin{figure}[th]
  \centering
  \includegraphics[width=0.48
\textwidth]{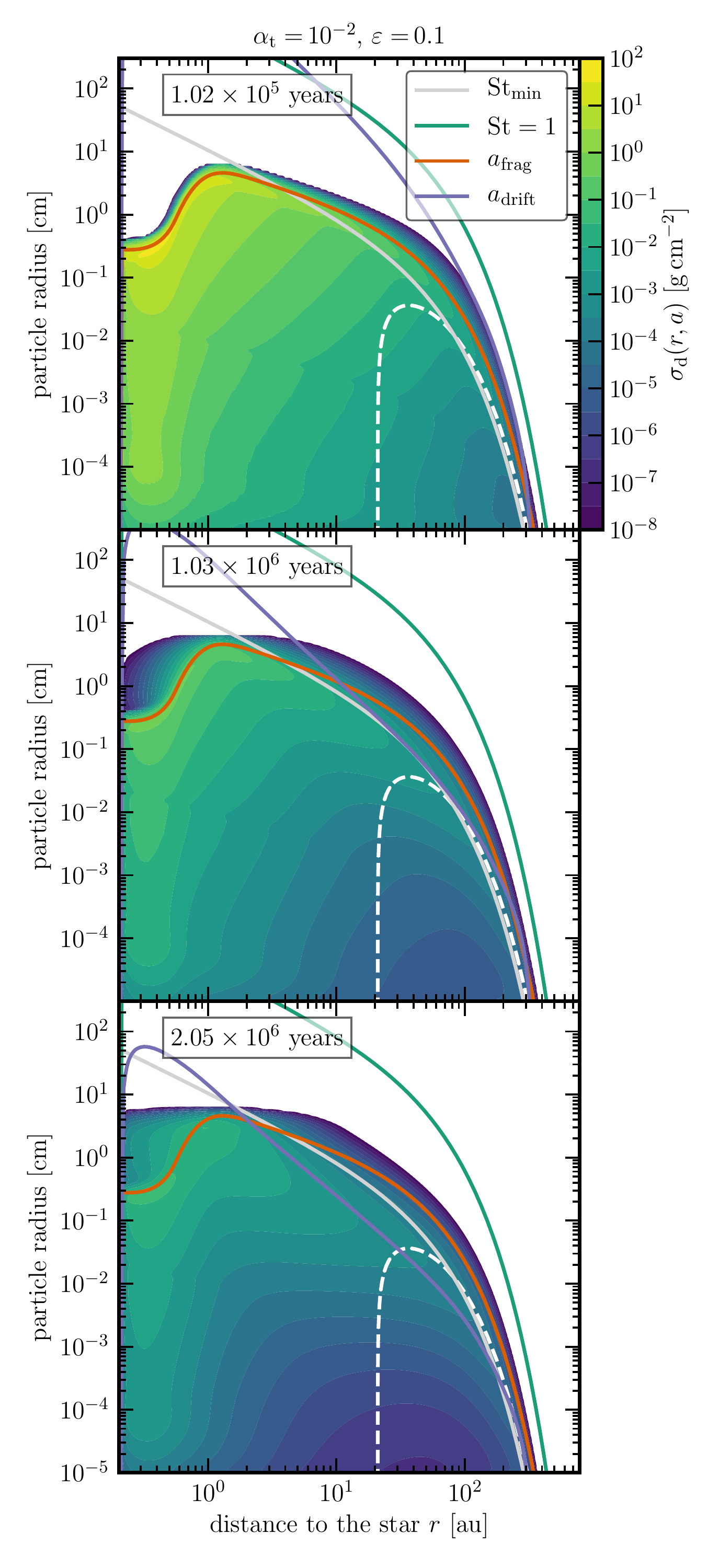}
  \caption
  {
  Same as Fig.~\ref{fig:s01_e01_r35_M5_flux_St_4_5_6_nodiskev} but showing particle column density per size bin (color) as function of particle radius. Furthermore, instead of the snapshot at $10^4\,\yr$ we show $2\cdot10^6\,\yr$. The gray line shows the
  size above which particles can potentially contribute to planetesimal formation.
   }
  \label{fig:s01_e01_a1e-2_r35_M5_dens_size_5_6_2e6_nodiskev}
\end{figure}

\begin{figure}[th]
  \centering
  \includegraphics[width=0.48\textwidth]{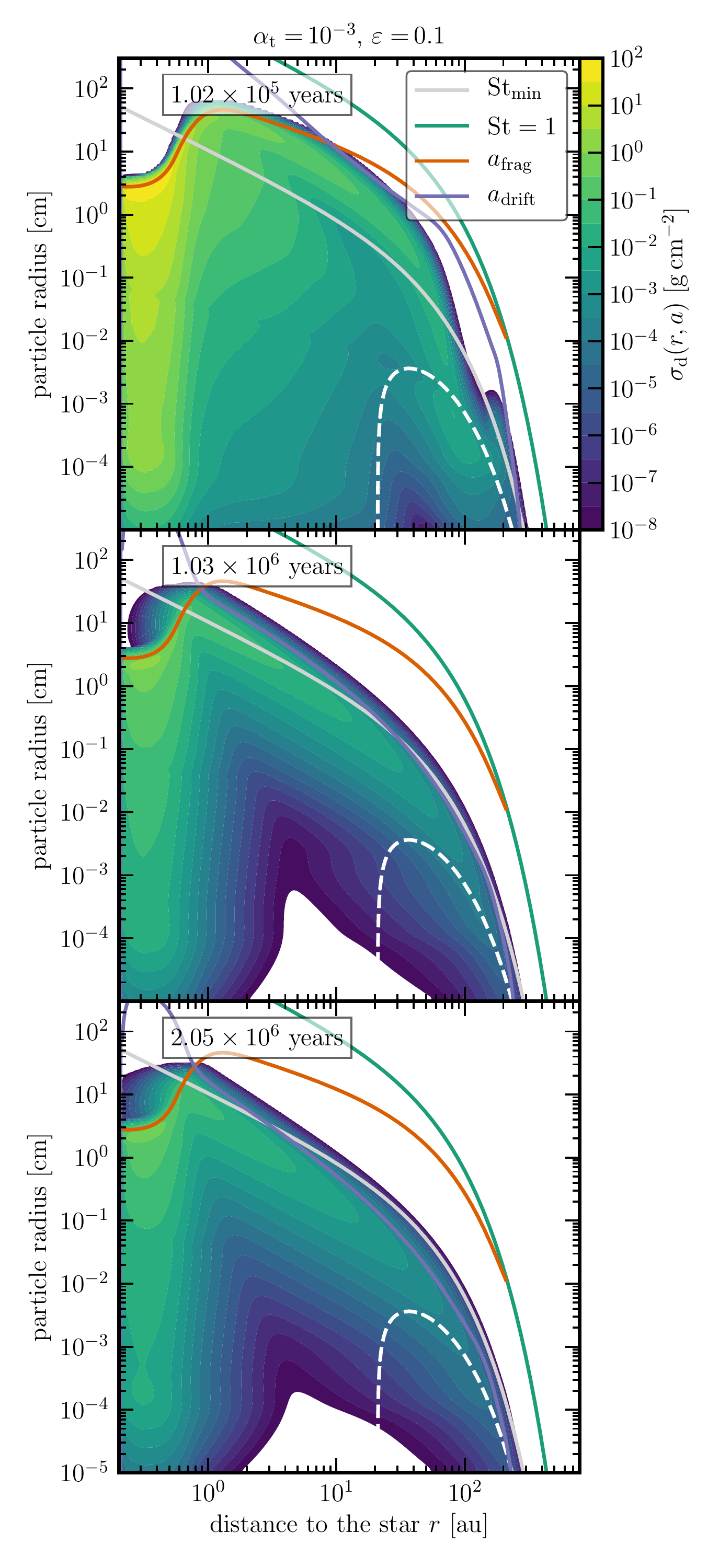}
  \caption
  {
  Same as Fig.~\ref{fig:s02_e01_r35_M5_flux_St_4_5_6_nodiskev} but showing particle column density per size bin (color) as function of particle radius. 
  Furthermore, instead of the snapshot at $10^4\,\yr$ we show $2\cdot10^6\,\yr$. The gray line shows the
  size above which particles can potentially contribute to planetesimal formation.
   }
  \label{fig:s02_e01_a1e-3_r35_M5_dens_size_5_6_2e6_nodiskev}
\end{figure}

Physical and numerical diffusion
mix particles radially, 
which can lead to particles larger than the growth barriers. 
They can remain there for a long time if the particle density is low enough to keep collision rates small, see figures~\ref{fig:s01_e01_a1e-2_r35_M5_dens_size_5_6_2e6_nodiskev} and \ref{fig:s02_e01_a1e-3_r35_M5_dens_size_5_6_2e6_nodiskev}. Converting particle radius into Stokes number, this effect looks more extreme in the outer disk (figures~\ref{fig:s01_e01_r35_M5_flux_St_4_5_6_nodiskev} and \ref{fig:s02_e01_r35_M5_flux_St_4_5_6_nodiskev}) because the gas density decreases with disk radius.

\section{More detailed derivation of the planetesimal formation condition}\label{cond_detail}
% Condition \eqref{eq:cond_ptes} assumes that the accumulation of pebbles in pressure bumps leads to Hill density on the smallest scales of planetesimal formation. Since the case where the model assumes an artificial turbulence in form of $\aturb$ within a fluid dynamical simulation leading to planetesimal formation was never studied, we can only guess, what the criterion(s) for planetesimal formation in this specific case would be. 
% critical pebble-to-gas ratios of $1$ in the midplane, triggering streaming instability \citep{YoudinGoodman2005}. In fact this would 
% be represented by the equation
% \begin{align}
% \begin{aligned}
%   1\leq\frac{\rho_\mathrm{peb,0}}{\rho_\mathrm{g,0}}\approx &
%   \sum_{\stokes}\frac{\Sigma_\mathrm{peb}(t_0,\stokes)}{\Sigma_\mathrm{g}(t_0)}\sqrt{1+\frac{\stokes}{\aturb}}
%   \\
%   &+\frac{\varepsilon_\mathrm{trap}}{\Sigma_\mathrm{g}(t_0)\ell_r\ell_\varphi/(2\pi r)}
%   \sum_{\stokes}\sqrt{1+\frac{\stokes}{\aturb}}
%   \\
%   &\hphantom{+}\underbrace{\int_{t_1}^{t}\Sigma_\mathrm{peb}(t,\stokes)\left|{v_\mathrm{peb}(t,\stokes)}\right|\mathrm{d}t}_{\approx\Sigma_\mathrm{peb}(t_0,\stokes)\left|{v_\mathrm{peb}(t_0,\stokes)}\right|\tau}.
%   \label{eq:condnew}
% %\mathrm{St_{min}}\leq\mathrm{St}\leq\mathrm{St_{max}}
% \end{aligned}
% \end{align}
In order to be able to form planetesimals, there exists a critical cloud diameter $\lc$ that has to be reached in order for the cloud to be 
able to contract while an underlying turbulent particle diffusion is trying to dissolve
it \citep[his section 3.3]{andydiss}. This critical length scale can be derived by equating the diffusion and collapse timescale leading to
\begin{align}
    \lc=\frac{2}{3}\sqrt{\frac{\delta}{\stokes}}\hscale.
\end{align}
% which is on the same order of magnitude as $\hscdust$. 
Here, $\delta$ is the dimensionless diffusion parameter (diffusion constant over $\sound\hscale$) which acts on the scale of $\lc$. 
% In order to be able to form planetesimals, 
Particles have to concentrate and reach Hill density $\rho_\mathrm{Hill}$ in a volume
$\lc^3$. \Ie, under self-gravity particles will settle to an effective scale height $\ell_z=\lc$ in the 
$z$-direction and concentrate azimuthally out of a full $2\pi r$ ring to a length of $\ell_\varphi=\lc$. In the radial direction 
the particles that have been in the trap structure of radial extent $d_r$ will be concentrated within the 
length $\ell_r=\lc$, and the particles that would drift the radial distance 
$\left|{v_\mathrm{peb}(t_0,\stokes)}\right|\tau$ during the time $\tau$ are also concentrated within $\ell_r=\lc$.
We add the factor $4\pi/3$ because these particles are more likely to concentrate into a spherical shape than a cubic one.
The condition for planetesimal formation then reads
\begin{align}
\begin{aligned}
  \rho_\mathrm{Hill}&\leq\rho_\mathrm{peb,l_c}
  \\
  &\approx
  \frac{\peff}{(4\pi/3)\ell_z\ell_r\ell_\varphi/(2\pi r)}
  \\
  &\hphantom{\approx}
  \sum_{\stmin\leq\stokes\leq\stmax}
  \Bigg(
    \int_{r-d_r/2}^{r+d_r/2}\Sigma_\mathrm{peb}(t_1,\stokes)\dif{r}
    \\
    &\hphantom{\approx}+
    \underbrace{\int_{t_1}^{t}\Sigma_\mathrm{peb}(t,\stokes)\left|{v_\mathrm{peb}(t,\stokes)}\right|\mathrm{d}t}_{\approx\Sigma_\mathrm{peb}(t_0,\stokes)\left|{v_\mathrm{peb}(t_0,\stokes)}\right|\tau}
  \Bigg).
  \label{eq:condnew}
%\mathrm{St_{min}}\leq\mathrm{St}\leq\mathrm{St_{max}}
\end{aligned}
\end{align}
The second term will mostly
dominate the first term. The expression 
\begin{align}
    \frac{4\pi}{3}\lc^3\rho_\mathrm{Hill}=\mptes
\end{align}
gives the resulting planetesimal mass \citep{KlahrSchreiber2015,andydiss} and, by neglecting the first 
term, we obtain the same condition as in Eq.~\eqref{eq:cond_ptes}.

% Here the $\stokes$ sum goes over all $\stmin\leq\stokes\leq\stmax$ particles, $\varepsilon_\mathrm{trap}$ is the pure 
% efficiency of trapping (without planetesimal formation), and $\ell_{r/\varphi}$ are the concentration factors in $r$ and 
% $\varphi$ direction. This can be condensed in one effective \emph{trapping and concentration efficiency parameter} 
% \begin{align}
%   \varepsilon_\mathrm{tc}:= \frac{\varepsilon_\mathrm{trap}2\pi r\hscale(r)}{\ell_{r}\ell_{\varphi}},
% \end{align}
% which can reach values larger than $1$. For an axisymmetric zonal flow one would have $\ell_{\varphi}=2\pi r$ but 
% vortices can concentrate material in a much smaller azimuthal extent.  According to \cite{surville2015}, the pure gas structures of vortices typically have a radial extent of $1-3\,\hscale$ and azimuthally $4-10$ times radial. 
The timescale over which the concentration will occur is assumed to be the lifetime of these structures, $\tau=\tlife$. 
% The square root factor arises due to the assumption of midplane values \eqref{eq:rhozerogas} and particle scale heights 
% \eqref{eq:hscdust} caused by vertical equilibrium of diffusion and settling.
Since we need a local criterion to make the model work, we have to assume that the particle flux does not change significantly over $\tau$, or if it does, that it will still lead to the same result of either
reaching or not reaching Hill density.
%(non-)critical pebble-to-gas ratios.

% \cite{johansen2009} have shown that ratios of $\dustsurf/\surfgas>(2-3)\cdot10^{-2}$ are enough to trigger streaming instability. But this also depends on 
% $\aturb$. Analytical calculations by \cite{auffinger2017} show that the classical streaming instability 
% in the $r$-$z$ (radial-vertical) plane prefer pressure bumps to be active. But there exists also streaming 
% instability in the $r$-$\varphi$ (radial-azimuthal) plane \citep{SchreiberKlahr2018}.

\end{appendices} 
\bibliography{bibliography.bib}

\begin{thebibliography}{134}
\expandafter\ifx\csname natexlab\endcsname\relax\def\natexlab#1{#1}\fi

\bibitem[{Abod {et~al.}(2018)Abod, Simon, Li, Armitage, Youdin, \&
  Kretke}]{abod2018}
Abod, C.~P., Simon, J.~B., Li, R., Armitage, P.~J., Youdin, A.~N., \& Kretke,
  K.~A. 2018, arXiv preprint arXiv:1810.10018

\bibitem[{Adachi {et~al.}(1976)Adachi, Hayashi, \& Nakazawa}]{adachi1976}
Adachi, I., Hayashi, C., \& Nakazawa, K. 1976, Progress of Theoretical Physics,
  56, 1756

\bibitem[{Andrews {et~al.}(2018)Andrews, Huang, P{\'e}rez, Isella, Dullemond,
  Kurtovic, Guzm{\'a}n, Carpenter, Wilner, Zhang, {et~al.}}]{andrews2018}
Andrews, S.~M. {et~al.} 2018, \apjl, 869, L41

\bibitem[{Andrews {et~al.}(2010)Andrews, Wilner, Hughes, Qi, \&
  Dullemond}]{andrews2010}
Andrews, S.~M., Wilner, D., Hughes, A., Qi, C., \& Dullemond, C. 2010, \apj,
  723, 1241

\bibitem[{Arlt \& Urpin(2004)}]{arlt2004}
Arlt, R., \& Urpin, V. 2004, \aap, 426, 755

\bibitem[{Auffinger \& Laibe(2017)}]{auffinger2017}
Auffinger, J., \& Laibe, G. 2017, \mnras, 473, 796

\bibitem[{Bai \& Stone(2010)}]{BaiStone2010}
Bai, X.-N., \& Stone, J.~M. 2010, \apj, 722, 1437

\bibitem[{Bai \& Stone(2014)}]{BaiStone2014}
---. 2014, \apj, 796, 31

\bibitem[{{Balbus} \& {Hawley}(1991)}]{BalbusHawley1991}
{Balbus}, S.~A., \& {Hawley}, J.~F. 1991, \apj, 376, 214

\bibitem[{Barker \& Latter(2015)}]{barker2015}
Barker, A.~J., \& Latter, H.~N. 2015, Mon. Not. R. Astron. Soc., 450, 21

\bibitem[{Barrow(1981)}]{barrow1981}
Barrow, J.~D. 1981, Journal of physics A: Mathematical and General, 14, 729

\bibitem[{Beitz {et~al.}(2011)Beitz, G{\"u}ttler, Blum, Meisner, Teiser, \&
  Wurm}]{Beitz2011}
Beitz, E., G{\"u}ttler, C., Blum, J., Meisner, T., Teiser, J., \& Wurm, G.
  2011, \apj, 736, 34

\bibitem[{{B{\'e}thune} {et~al.}(2016){B{\'e}thune}, {Lesur}, \&
  {Ferreira}}]{bethune2016}
{B{\'e}thune}, W., {Lesur}, G., \& {Ferreira}, J. 2016, \aap, 589, A87

\bibitem[{B{\'e}thune {et~al.}(2017)B{\'e}thune, Lesur, \&
  Ferreira}]{bethune2017}
B{\'e}thune, W., Lesur, G., \& Ferreira, J. 2017, \aap, 600, A75

\bibitem[{Birnstiel {et~al.}(2010)Birnstiel, Dullemond, \&
  Brauer}]{birnstiel2010}
Birnstiel, T., Dullemond, C.~P., \& Brauer, F. 2010, \aap, 513, A79

\bibitem[{Birnstiel {et~al.}(2016)Birnstiel, Fang, \& Johansen}]{birnstiel2016}
Birnstiel, T., Fang, M., \& Johansen, A. 2016, Space Science Reviews, 205, 41

\bibitem[{{Birnstiel} {et~al.}(2012){Birnstiel}, {Klahr}, \&
  {Ercolano}}]{Birnstiel2012}
{Birnstiel}, T., {Klahr}, H., \& {Ercolano}, B. 2012, \aap, 539, A148

\bibitem[{Blum {et~al.}(2017)Blum, Gundlach, Krause, Fulle, Johansen, Agarwal,
  von Borstel, Shi, Hu, Bentley, {et~al.}}]{blum2017}
Blum, J. {et~al.} 2017, \mnras, 469, S755

\bibitem[{Blum \& M{\"u}nch(1993)}]{BlumMuench1993}
Blum, J., \& M{\"u}nch, M. 1993, Icarus, 106, 151

\bibitem[{Blum \& Wurm(2008)}]{BlumWurm2008}
Blum, J., \& Wurm, G. 2008, Annu. Rev. Astron. Astrophys., 46, 21

\bibitem[{Bottke~Jr {et~al.}(2005)Bottke~Jr, Durda, Nesvorn{\`y}, Jedicke,
  Morbidelli, Vokrouhlick{\`y}, \& Levison}]{bottke2005}
Bottke~Jr, W.~F., Durda, D.~D., Nesvorn{\`y}, D., Jedicke, R., Morbidelli, A.,
  Vokrouhlick{\`y}, D., \& Levison, H. 2005, Icarus, 175, 111

\bibitem[{Brauer {et~al.}(2008{\natexlab{a}})Brauer, Dullemond, \&
  Henning}]{brauer2008coagulation}
Brauer, F., Dullemond, C.~P., \& Henning, T. 2008{\natexlab{a}}, \aap, 480, 859

\bibitem[{Brauer {et~al.}(2008{\natexlab{b}})Brauer, Henning, \&
  Dullemond}]{brauer2008planetesimal}
Brauer, F., Henning, T., \& Dullemond, C.~P. 2008{\natexlab{b}}, \aap, 487, L1

\bibitem[{Carrera {et~al.}(2017)Carrera, Gorti, Johansen, \&
  Davies}]{carrera2017}
Carrera, D., Gorti, U., Johansen, A., \& Davies, M.~B. 2017, \apj, 839, 16

\bibitem[{{Carrera} {et~al.}(2015){Carrera}, {Johansen}, \&
  {Davies}}]{Carrera2015}
{Carrera}, D., {Johansen}, A., \& {Davies}, M.~B. 2015, \aap, 579, A43

\bibitem[{Carry(2012)}]{carry2012}
Carry, B. 2012, Planetary and Space Science, 73, 98

\bibitem[{Chiang \& Goldreich(1997)}]{chiang1997}
Chiang, E., \& Goldreich, P. 1997, \apj, 490, 368

\bibitem[{Cuzzi {et~al.}(2001)Cuzzi, Hogan, Paque, \& Dobrovolskis}]{cuzzi2001}
Cuzzi, J.~N., Hogan, R.~C., Paque, J.~M., \& Dobrovolskis, A.~R. 2001, \apj,
  546, 496

\bibitem[{Cuzzi {et~al.}(2008)Cuzzi, Hogan, \& Shariff}]{cuzzi2008}
Cuzzi, J.~N., Hogan, R.~C., \& Shariff, K. 2008, \apj, 687, 1432

\bibitem[{Delbo {et~al.}(2017)Delbo, Walsh, Bolin, Avdellidou, \&
  Morbidelli}]{delbo2017}
Delbo, M., Walsh, K., Bolin, B., Avdellidou, C., \& Morbidelli, A. 2017,
  Science, 357, 1026

\bibitem[{{Dittrich} {et~al.}(2013){Dittrich}, {Klahr}, \&
  {Johansen}}]{Dittrich2013}
{Dittrich}, K., {Klahr}, H., \& {Johansen}, A. 2013, \apj, 763, 117

\bibitem[{Dr{\k{a}}{\.z}kowska \& Alibert(2017)}]{drazkowska2017}
Dr{\k{a}}{\.z}kowska, J., \& Alibert, Y. 2017, \aap, 608, A92

\bibitem[{Dr{\k{a}}{\.z}kowska {et~al.}(2016)Dr{\k{a}}{\.z}kowska, Alibert, \&
  Moore}]{drazkowska2016}
Dr{\k{a}}{\.z}kowska, J., Alibert, Y., \& Moore, B. 2016, \aap, 594, A105

\bibitem[{Dr{\k{a}}{\.z}kowska \& Dullemond(2018)}]{drazkowska2018}
Dr{\k{a}}{\.z}kowska, J., \& Dullemond, C.~P. 2018, \aap, 614, A62

\bibitem[{Dr{\k{a}}{\.z}kowska {et~al.}(2013)Dr{\k{a}}{\.z}kowska, Windmark, \&
  Dullemond}]{drazkowska2013}
Dr{\k{a}}{\.z}kowska, J., Windmark, F., \& Dullemond, C. 2013, \aap, 556, A37

\bibitem[{Dr{\k{a}}{\.z}kowska
  {et~al.}(2014{\natexlab{a}})Dr{\k{a}}{\.z}kowska, Windmark, \&
  Dullemond}]{drazkowska2014}
Dr{\k{a}}{\.z}kowska, J., Windmark, F., \& Dullemond, C.~P. 2014{\natexlab{a}},
  \aap, 567, A38

\bibitem[{Dr{\k{a}}{\.z}kowska
  {et~al.}(2014{\natexlab{b}})Dr{\k{a}}{\.z}kowska, Windmark, \&
  Okuzumi}]{drazkowska2014proceeding}
Dr{\k{a}}{\.z}kowska, J., Windmark, F., \& Okuzumi, S. 2014{\natexlab{b}},
  Proceedings of the International Astronomical Union, 9, 208

\bibitem[{Dubrulle {et~al.}(1995)Dubrulle, Morfill, \& Sterzik}]{Dubrulle1995}
Dubrulle, B., Morfill, G., \& Sterzik, M. 1995, Icarus, 114, 237

\bibitem[{Dullemond \& Dominik(2005)}]{DullemondDominik2005}
Dullemond, C., \& Dominik, C. 2005, \aap, 434, 971

\bibitem[{Dullemond {et~al.}(2018)Dullemond, Birnstiel, Huang, Kurtovic,
  Andrews, Guzm{\'a}n, P{\'e}rez, Isella, Zhu, Benisty,
  {et~al.}}]{dullemond2018}
Dullemond, C.~P. {et~al.} 2018, \apjl, 869, L46

\bibitem[{Epstein(1924)}]{epstein1924}
Epstein, P.~S. 1924, Physical Review, 23, 710

\bibitem[{Estrada {et~al.}(2016)Estrada, Cuzzi, \& Morgan}]{estrada2016}
Estrada, P.~R., Cuzzi, J.~N., \& Morgan, D.~A. 2016, \apj, 818, 200

\bibitem[{Fraser \& Kavelaars(2008)}]{FraserKavelaars2008}
Fraser, W.~C., \& Kavelaars, J. 2008, The Astronomical Journal, 137, 72

\bibitem[{Fu {et~al.}(2014)Fu, Li, Lubow, Li, \& Liang}]{fu2014}
Fu, W., Li, H., Lubow, S., Li, S., \& Liang, E. 2014, \apjl, 795, L39

\bibitem[{Fuentes \& Holman(2008)}]{FuentesHolman2008}
Fuentes, C.~I., \& Holman, M.~J. 2008, \apj, 136, 83

\bibitem[{{Goldreich} \& {Ward}(1973)}]{GoldreichWard1973}
{Goldreich}, P., \& {Ward}, W.~R. 1973, \apj, 183, 1051

\bibitem[{Gundlach \& Blum(2014)}]{GundlachBlum2014}
Gundlach, B., \& Blum, J. 2014, \apj, 798, 34

\bibitem[{G{\"u}ttler {et~al.}(2010)G{\"u}ttler, Blum, Zsom, Ormel, \&
  Dullemond}]{guttler2010}
G{\"u}ttler, C., Blum, J., Zsom, A., Ormel, C.~W., \& Dullemond, C.~P. 2010,
  \aap, 513, A56

\bibitem[{Hayashi(1981)}]{Hayashi1981}
Hayashi, C. 1981, Progress of Theoretical Physics Supplement, 70

\bibitem[{Huang {et~al.}(2018)Huang, Andrews, Dullemond, Isella, P{\'e}rez,
  Guzm{\'a}n, {\"O}berg, Zhu, Zhang, Bai, {et~al.}}]{huang2018}
Huang, J. {et~al.} 2018, \apjl, 869, L42

\bibitem[{Husmann(2017)}]{HusmannDiss}
Husmann, T. 2017, PhD thesis, University of Duisburg-Essen

\bibitem[{Jewitt {et~al.}(2000)Jewitt, Trujillo, \& Luu}]{Jewitt2000}
Jewitt, D.~C., Trujillo, C.~A., \& Luu, J.~X. 2000, The Astronomical Journal,
  120, 1140

\bibitem[{{Johansen} {et~al.}(2006){Johansen}, {Henning}, \&
  {Klahr}}]{JohansenHenningKlahr2006}
{Johansen}, A., {Henning}, T., \& {Klahr}, H. 2006, \apj, 643, 1219

\bibitem[{Johansen {et~al.}(2006)Johansen, Klahr, \& Henning}]{Johansen2006}
Johansen, A., Klahr, H., \& Henning, T. 2006, \apj, 636, 1121

\bibitem[{Johansen {et~al.}(2011)Johansen, Klahr, \& Henning}]{johansen2011}
---. 2011, \aap, 529, A62

\bibitem[{{Johansen} {et~al.}(2015){Johansen}, {Mac Low}, {Lacerda}, \&
  {Bizzarro}}]{Johansen2015}
{Johansen}, A., {Mac Low}, M.-M., {Lacerda}, P., \& {Bizzarro}, M. 2015,
  Science Advances, 1, e1500109

\bibitem[{{Johansen} {et~al.}(2007){Johansen}, {Oishi}, {Mac Low}, {Klahr},
  {Henning}, \& {Youdin}}]{johansen2007nat}
{Johansen}, A., {Oishi}, J.~S., {Mac Low}, M.-M., {Klahr}, H., {Henning}, T.,
  \& {Youdin}, A. 2007, \nat, 448, 1022

\bibitem[{Johansen {et~al.}(2009{\natexlab{a}})Johansen, Youdin, \&
  Klahr}]{johansen2009zonal}
Johansen, A., Youdin, A., \& Klahr, H. 2009{\natexlab{a}}, \apj, 697, 1269

\bibitem[{Johansen {et~al.}(2009{\natexlab{b}})Johansen, {Youdin}, \& {Mac
  Low}}]{johansen2009}
Johansen, A., {Youdin}, A., \& {Mac Low}, M.-M. 2009{\natexlab{b}}, \apjl, 704,
  L75

\bibitem[{Johansen \& Youdin(2007)}]{Johansen2007}
Johansen, A., \& Youdin, A.~N. 2007, \apj, 662, 627

\bibitem[{Johansen {et~al.}(2012)Johansen, Youdin, \& Lithwick}]{johansen2012}
Johansen, A., Youdin, A.~N., \& Lithwick, Y. 2012, \aap, 537, A125

\bibitem[{Kataoka {et~al.}(2013)Kataoka, Tanaka, Okuzumi, \&
  Wada}]{kataoka2013}
Kataoka, A., Tanaka, H., Okuzumi, S., \& Wada, K. 2013, \aap, 557, L4

\bibitem[{Klahr \& Bodenheimer(2006)}]{KlahrBodenheimer2006}
Klahr, H., \& Bodenheimer, P. 2006, \apj, 639, 432

\bibitem[{Klahr \& Hubbard(2014)}]{KlahrHubbard2014}
Klahr, H., \& Hubbard, A. 2014, \apj, 788, 21

\bibitem[{Klahr {et~al.}(2018)Klahr, Pfeil, \& Schreiber}]{klahr2018}
Klahr, H., Pfeil, T., \& Schreiber, A. 2018, Handbook of Exoplanets, 2251

\bibitem[{Klahr \& Schreiber(2015)}]{KlahrSchreiber2015}
Klahr, H., \& Schreiber, A. 2015, Proceedings of the International Astronomical
  Union, 10, 1

\bibitem[{Klahr \& Bodenheimer(2003)}]{KlahrBodenheimer2003}
Klahr, H.~H., \& Bodenheimer, P. 2003, \apj, 582, 869

\bibitem[{Klessen {et~al.}(2005)Klessen, Spaans, \& Jappsen}]{klessen2005}
Klessen, R.~S., Spaans, M., \& Jappsen, A.-K. 2005, Proceedings of the
  International Astronomical Union, 1, 337

\bibitem[{Kobayashi {et~al.}(2016)Kobayashi, Tanaka, \&
  Okuzumi}]{kobayashi2016}
Kobayashi, H., Tanaka, H., \& Okuzumi, S. 2016, \apj, 817, 105

\bibitem[{Krijt {et~al.}(2016{\natexlab{a}})Krijt, Ciesla, \&
  Bergin}]{krijt2016b}
Krijt, S., Ciesla, F.~J., \& Bergin, E.~A. 2016{\natexlab{a}}, \apj, 833, 285

\bibitem[{Krijt {et~al.}(2015)Krijt, Ormel, Dominik, \& Tielens}]{krijt2015}
Krijt, S., Ormel, C.~W., Dominik, C., \& Tielens, A.~G. 2015, \aap, 574, A83

\bibitem[{Krijt {et~al.}(2016{\natexlab{b}})Krijt, Ormel, Dominik, \&
  Tielens}]{krijt2016a}
---. 2016{\natexlab{b}}, \aap, 586, A20

\bibitem[{Latter \& Papaloizou(2017)}]{latter2017}
Latter, H.~N., \& Papaloizou, J. 2017, \mnras, 474, 3110

\bibitem[{Lee(2000)}]{lee2000}
Lee, M.~H. 2000, Icarus, 143, 74

\bibitem[{Leinhardt \& Stewart(2009)}]{leinhardt2009}
Leinhardt, Z.~M., \& Stewart, S.~T. 2009, Icarus, 199, 542

\bibitem[{Lesur \& Papaloizou(2010)}]{lesur2010}
Lesur, G., \& Papaloizou, J.~C. 2010, \aap, 513, A60

\bibitem[{Lodders(2003)}]{lodders2003}
Lodders, K. 2003, \apj, 591, 1220

\bibitem[{L{\"u}st(1952)}]{Luest1952}
L{\"u}st, R. 1952, Zeitschrift Naturforschung Teil A, 7, 87

\bibitem[{Lynden-Bell \& Pringle(1974)}]{lbp1974}
Lynden-Bell, D., \& Pringle, J. 1974, \mnras, 168, 603

\bibitem[{Lyra(2014)}]{lyra2014}
Lyra, W. 2014, \apj, 789, 77

\bibitem[{Lyra \& Klahr(2011)}]{LyraKlahr2011}
Lyra, W., \& Klahr, H. 2011, \aap, 527, A138

\bibitem[{Lyra \& Umurhan(2018)}]{LyraUmurhan2018}
Lyra, W., \& Umurhan, O. 2018, arXiv preprint arXiv:1808.08681

\bibitem[{Manger \& Klahr(2018)}]{MangerKlahr2018}
Manger, N., \& Klahr, H. 2018, \mnras, 480, 2125

\bibitem[{Mathis {et~al.}(1977)Mathis, Rumpl, \& Nordsieck}]{mathis1977}
Mathis, J.~S., Rumpl, W., \& Nordsieck, K.~H. 1977, \apj, 217, 425

\bibitem[{Min {et~al.}(2011)Min, Dullemond, Kama, \& Dominik}]{min2011}
Min, M., Dullemond, C., Kama, M., \& Dominik, C. 2011, Icarus, 212, 416

\bibitem[{Morbidelli {et~al.}(2009)Morbidelli, Bottke, Nesvorn{\`y}, \&
  Levison}]{morbidelli2009}
Morbidelli, A., Bottke, W.~F., Nesvorn{\`y}, D., \& Levison, H.~F. 2009,
  Icarus, 204, 558

\bibitem[{{Nakagawa} {et~al.}(1986){Nakagawa}, {Sekiya}, \&
  {Hayashi}}]{Nakagawa1986}
{Nakagawa}, Y., {Sekiya}, M., \& {Hayashi}, C. 1986, \icarus, 67, 375

\bibitem[{Nelson {et~al.}(2013)Nelson, Gressel, \& Umurhan}]{nelson2013}
Nelson, R.~P., Gressel, O., \& Umurhan, O.~M. 2013, \mnras, 435, 2610

\bibitem[{Okuzumi(2009)}]{okuzumi2009}
Okuzumi, S. 2009, \apj, 698, 1122

\bibitem[{Okuzumi {et~al.}(2012)Okuzumi, Tanaka, Kobayashi, \&
  Wada}]{okuzumi2012}
Okuzumi, S., Tanaka, H., Kobayashi, H., \& Wada, K. 2012, \apj, 752, 106

\bibitem[{Ormel \& Cuzzi(2007)}]{OrmelCuzzi2007}
Ormel, C., \& Cuzzi, J. 2007, \aap, 466, 413

\bibitem[{Ormel \& Klahr(2010)}]{OrmelKlahr2010}
Ormel, C., \& Klahr, H. 2010, \aap, 520, A43

\bibitem[{Owen {et~al.}(2012)Owen, Clarke, \& Ercolano}]{owen2012}
Owen, J.~E., Clarke, C.~J., \& Ercolano, B. 2012, \mnras, 422, 1880

\bibitem[{Papaloizou \& Lin(1995)}]{PapaloizouLin1995}
Papaloizou, J., \& Lin, D. 1995, \araa, 33, 505

\bibitem[{Petersen {et~al.}(2007{\natexlab{a}})Petersen, Julien, \&
  Stewart}]{petersen2007a}
Petersen, M.~R., Julien, K., \& Stewart, G.~R. 2007{\natexlab{a}}, \apj, 658,
  1236

\bibitem[{Petersen {et~al.}(2007{\natexlab{b}})Petersen, Stewart, \&
  Julien}]{petersen2007b}
Petersen, M.~R., Stewart, G.~R., \& Julien, K. 2007{\natexlab{b}}, \apj, 658,
  1252

\bibitem[{Pfeil \& Klahr(2019)}]{PfeilKlahr2018}
Pfeil, T., \& Klahr, H. 2019, \apj, 871, 150

\bibitem[{{Pringle}(1981)}]{Pringle1981}
{Pringle}, J.~E. 1981, \araa, 19, 137

\bibitem[{{Raettig} {et~al.}(2015){Raettig}, {Klahr}, \& {Lyra}}]{Raettig2015}
{Raettig}, N., {Klahr}, H., \& {Lyra}, W. 2015, \apj, 804, 35

\bibitem[{Raettig {et~al.}(2013)Raettig, Lyra, \& Klahr}]{raettig2013}
Raettig, N., Lyra, W., \& Klahr, H. 2013, \apj, 765, 115

\bibitem[{{Safronov}(1969)}]{Safronov1969}
{Safronov}, V.~S. 1969, {Evoliutsiia Doplanetnogo Oblaka.} (English transl.:
  Evolution of the protoplanetary cloud and formation of Earth and the planets,
  NASA Tech. Transl. F-677, Jerusalem: Israel Sci. Transl. 1972)

\bibitem[{Savage \& Jenkins(1972)}]{savage1972}
Savage, B.~D., \& Jenkins, E.~B. 1972, \apj, 172, 491

\bibitem[{Schreiber(2018)}]{andydiss}
Schreiber, A. 2018, PhD thesis, Ruperto-Carola University Heidelberg

\bibitem[{Schreiber \& Klahr(2018)}]{SchreiberKlahr2018}
Schreiber, A., \& Klahr, H. 2018, \apj, 861, 47

\bibitem[{Schumann(1940)}]{Schumann1940}
Schumann, T. 1940, Quarterly Journal of the Royal Meteorological Society, 66,
  195

\bibitem[{Sekiya(1998)}]{sekiya1998}
Sekiya, M. 1998, Icarus, 133, 298

\bibitem[{{Shakura} \& {Sunyaev}(1973)}]{Shakura1973}
{Shakura}, N.~I., \& {Sunyaev}, R.~A. 1973, \aap, 24, 337

\bibitem[{Sheppard \& Trujillo(2010)}]{SheppardTrujillo2010}
Sheppard, S.~S., \& Trujillo, C.~A. 2010, \apjl, 723, L233

\bibitem[{Simon {et~al.}(2016)Simon, Armitage, Li, \& Youdin}]{simon2016}
Simon, J.~B., Armitage, P.~J., Li, R., \& Youdin, A.~N. 2016, \apj, 822, 55

\bibitem[{Squire \& Hopkins(2018{\natexlab{a}})}]{SquireHopkins2018b}
Squire, J., \& Hopkins, P.~F. 2018{\natexlab{a}}, \mnras, 477, 5011

\bibitem[{Squire \& Hopkins(2018{\natexlab{b}})}]{SquireHopkins2018a}
---. 2018{\natexlab{b}}, \apjl, 856, L15

\bibitem[{Stammler {et~al.}(2017)Stammler, Birnstiel, Pani{\'c}, Dullemond, \&
  Dominik}]{stammler2017}
Stammler, S.~M., Birnstiel, T., Pani{\'c}, O., Dullemond, C.~P., \& Dominik, C.
  2017, \aap, 600, A140

\bibitem[{Stokes(1851)}]{stokes1851}
Stokes, G.~G. 1851, On the effect of the internal friction of fluids on the
  motion of pendulums, Vol.~9 (Pitt Press Cambridge)

\bibitem[{Stoll \& Kley(2014)}]{StollKley2014}
Stoll, M.~H., \& Kley, W. 2014, \aap, 572, A77

\bibitem[{Stoll \& Kley(2016)}]{StollKley2016}
---. 2016, \aap, 594, A57

\bibitem[{Takeuchi \& Lin(2002)}]{TakeuchiLin2002}
Takeuchi, T., \& Lin, D. 2002, \apj, 581, 1344

\bibitem[{Tanaka {et~al.}(2005)Tanaka, Himeno, \& Ida}]{tanaka2005}
Tanaka, H., Himeno, Y., \& Ida, S. 2005, \apj, 625, 414

\bibitem[{{Urpin} \& {Brandenburg}(1998)}]{urpin1998}
{Urpin}, V., \& {Brandenburg}, A. 1998, \mnras, 294, 399

\bibitem[{von Smoluchowski(1916)}]{Smoluchowski1916}
von Smoluchowski, M. 1916, Z. Phys., 17, 557

\bibitem[{Wada {et~al.}(2009)Wada, Tanaka, Suyama, Kimura, \&
  Yamamoto}]{wada2009}
Wada, K., Tanaka, H., Suyama, T., Kimura, H., \& Yamamoto, T. 2009, \apj, 702,
  1490

\bibitem[{{Weidenschilling}(1977)}]{weidenschilling1977}
{Weidenschilling}, S.~J. 1977, \mnras, 180, 57

\bibitem[{Weidenschilling(1977)}]{weidenschilling1977mmsn}
Weidenschilling, S.~J. 1977, Astrophysics and Space Science, 51, 153

\bibitem[{Weidenschilling(1980)}]{weidenschilling1980}
---. 1980, Icarus, 44, 172

\bibitem[{Weidenschilling(1995)}]{weidenschilling1995}
---. 1995, Icarus, 116, 433

\bibitem[{Weizs{\"a}cker(1948)}]{weizsacker1948}
Weizs{\"a}cker, C.~F. 1948, Zeitschrift f{\"u}r Naturforschung A, 3, 524

\bibitem[{Whipple(1972)}]{whipple1972}
Whipple, F.~L. 1972, in From plasma to planet, 211

\bibitem[{Windmark {et~al.}(2012)Windmark, Birnstiel, Ormel, \&
  Dullemond}]{windmark2012}
Windmark, F., Birnstiel, T., Ormel, C.~W., \& Dullemond, C.~P. 2012, \aap, 544,
  L16

\bibitem[{{Yang} {et~al.}(2017){Yang}, {Johansen}, \& {Carrera}}]{yang2017}
{Yang}, C.~C., {Johansen}, A., \& {Carrera}, D. 2017, \aap, 606, A80

\bibitem[{Yang {et~al.}(2018)Yang, Mac~Low, \& Johansen}]{yang2018}
Yang, C.-C., Mac~Low, M.-M., \& Johansen, A. 2018, \apj, 868, 27

\bibitem[{Youdin \& Goodman(2005)}]{YoudinGoodman2005}
Youdin, A.~N., \& Goodman, J. 2005, \apj, 620, 459

\bibitem[{Youdin \& Lithwick(2007)}]{YoudinLithwick2007}
Youdin, A.~N., \& Lithwick, Y. 2007, Icarus, 192, 588

\bibitem[{Youdin \& Shu(2002)}]{YoudinShu2002}
Youdin, A.~N., \& Shu, F.~H. 2002, \apj, 580, 494

\bibitem[{Zsom {et~al.}(2010)Zsom, Ormel, G{\"u}ttler, Blum, \&
  Dullemond}]{zsom2010}
Zsom, A., Ormel, C., G{\"u}ttler, C., Blum, J., \& Dullemond, C. 2010, \aap,
  513, A57

\bibitem[{Zvyagina {et~al.}(1974)Zvyagina, Pechernikova, \&
  Safronov}]{zvyagina1974}
Zvyagina, E., Pechernikova, G., \& Safronov, V. 1974, Soviet Astronomy, 17, 793

\end{thebibliography}

\end{document}